\documentclass[prd,twocolumn,amsmath,amsfonts,amssymb,showpacs,nofootinbib,preprintnumbers]{revtex4}
\usepackage{epsfig,graphicx,bm,color}
\begin{document}
\preprint{Phys.\ Rev.\ D \textbf{85}, 125027}
\title{Improved constraints on the primordial power spectrum at small scales from ultracompact minihalos}
\author{Torsten Bringmann}
\email{torsten.bringmann@desy.de}
\affiliation{{II.} Institute for Theoretical Physics, University of Hamburg, 
Luruper Chausse 149, DE-22761 Hamburg, Germany}
\author{Pat Scott}
\email{patscott@physics.mcgill.ca}
\affiliation{Department of Physics, McGill University, 
3600 rue University, Montr\'eal, QC, H3A 2T8, Canada}
\author{Yashar Akrami}
\email{yashar.akrami@astro.uio.no}
\affiliation{The Oskar Klein Centre for Cosmoparticle Physics, Department of Physics, Stockholm University, AlbaNova, SE-106 91 Stockholm, Sweden\\and\\Institute of Theoretical Astrophysics, University of Oslo, P.O. Box 1029 Blindern, N-0315 Oslo, Norway}


\begin{abstract}
For a Gaussian spectrum of primordial density fluctuations, ultracompact minihalos (UCMHs) of dark matter are expected to be produced in much greater abundance than, e.g., primordial black holes. Forming shortly after matter-radiation equality, these objects would develop very dense and spiky dark matter profiles. In the standard scenario where dark matter consists of thermally-produced, weakly-interacting massive particles, UCMHs could thus appear as highly luminous gamma-ray sources, or leave an imprint in the cosmic microwave background by changing the reionisation history of the Universe. 
We derive corresponding limits on the cosmic abundance of UCMHs at different epochs, and translate them into constraints on the primordial power spectrum.  We find the resulting constraints to be quite severe, especially at length scales much smaller than what can be directly probed by the cosmic microwave background or large-scale structure observations. We use our results to provide an updated compilation of the best available constraints on the power of density fluctuations on all scales, ranging from the present-day horizon to scales more than 20 orders of magnitude smaller.

\end{abstract}

\maketitle

\newcommand{\be}{\begin{equation}}
\newcommand{\ee}{\end{equation}}
\newcommand{\bea}{\begin{eqnarray}}
\newcommand{\eea}{\end{eqnarray}}

\hyphenation{}

\section{Introduction}

Structure formation in the Universe is thought to have proceeded in a bottom-up process, starting with the formation of smaller bound objects from   
 the gravitational collapse of small-amplitude, random initial density perturbations; in a second step, these smaller objects would then merge to larger ones.  In the simplest scenario \cite{Starobinsky:1982ee,Guth:1982ec,Bardeen:1983qw,Kodama:1985bj,Mukhanov:1990me,Riotto:2002yw} (see also Refs.~\cite{cosmology:Weinberg,cosmology:Mukhanov} for an introduction to the standard model of cosmology), a nearly scale-invariant (Harrison-Zel'dovich) \cite{Harrison:1969fb,Zeldovich:1972zz} Gaussian spectrum of perturbations is produced during inflation, and the bulk of perturbations are of sufficiently small amplitude at horizon entry ($\delta\sim10^{-5}$) that they do not collapse until well after matter-radiation equality.

None of these assumptions need be fulfilled, however.  Many inflationary models (and some alternatives) predict a departure from scale invariance (see e.g. Refs.~\cite{Kofman:1988xg,Starobinsky:1992ts,Chung:1999ve,Martin:2000xs,Kaloper:2003nv,Hunt:2004vt,Joy:2007na,Jain:2008dw,Ackerman:2010he}), or from Gaussian statistics (see e.g. Refs.~\cite{Bartolo:2004if,Chen:2010xka,Byrnes:2010em,Liguori:2010hx,Komatsu:2010hc} and references therein). In theories with more complicated cosmological histories, events such as phase transitions and late-time dynamics of scalar fields such as  string moduli, the inflaton or curvaton \cite{Schmid:1996qd,Adams97,Barriga01,Adams:2001vc,Ashoorioon:2008qr,Erickcek11}  might have lead to an injection of additional power on specific scales.  Regardless of when the primordial spectrum was produced, the presence of additional power on some scale may have led structures of a size corresponding to that scale to collapse far earlier than in the canonical scenario.  In severe cases, this collapse may even have occurred before matter-radiation equality \cite{Berezinsky:2010kq}.

The most extreme and best-studied example of such rapid gravitational collapse is that of primordial black holes (PBHs) \cite{PBHZeldovich,PBHHawking,CarrHawking}.  A PBH is expected to form when a perturbation enters the horizon with such a large amplitude ($\delta\gtrsim0.3-0.7$) that a substantial fraction of the horizon volume collapses directly to a black hole \cite{pbh_an,pbh_num}. Even very large density perturbations with $\delta\gtrsim1$ form PBHs, as has only recently been clarified \cite{PBHStefan}, rather than closing up upon themselves and forming separate universes, as originally argued by Hawking \cite{PBHHawking}.  PBHs have been proposed as possible dark matter (DM) candidates \cite{Ivanov:1994pa,Blais:2002nd,pbh2,Frampton:2010sw}, but the limits on their abundance are tight \cite{JGM09,Lacki:2010zf}, and the formation process lacks a natural mechanism for producing the observed cosmological abundance of DM.  Furthermore, 
due to the enormous amplitudes required for their formation,
a sizable PBH abundance can only be produced in scenarios that deviate rather strongly from the simple Harrison-Zel'dovich spectrum 
\cite{GreenLiddle,pbh1,pbh2,Josan:2010cj}.

Less severe departures, at the level of $\delta\sim10^{-3}$ during early radiation domination, can instead lead to the formation of so-called ultracompact primordial minihalos (UCMHs) \cite{Ricotti:2009bs,Berezinsky:2003vn,SS09}. 
Such a perturbation in the DM component will continue to grow until it eventually undergoes gravitational collapse before or very shortly after recombination. The mass of these objects is  proportional to the horizon mass at the time such a density perturbation enters the horizon, i.e.~it is uniquely determined by the wavelength of the perturbation mode.  In general, smaller-scale perturbations of a given amplitude will more easily produce UCMHs, as they enter the horizon earlier and thus give the over-density more time to grow.  Unlike in the PBH case, the lack of an event horizon means that the seed cannot retain radiation (which, at this early time, includes baryons), so baryonic matter and photons free-stream immediately back out of the over-density.  The seed thus consists entirely of cold dark matter (CDM), but it will later accrete both dark and (after recombination) baryonic matter.

The amplitudes of temperature inhomogeneities in the cosmic microwave background (CMB) \cite{cobe,wmap} allow us to infer the power of the primordial density fluctuations on scales that range from $\mathcal{O}(10^4\,$Mpc), almost the horizon size today, down to $\mathcal{O}(10\,$Mpc), roughly an order of magnitude smaller than the size of the sound horizon at the time of recombination.
The observed large-scale distribution of matter in the Universe also carries clues as to the amplitude of linear cosmological density perturbations, on  scales from about $10^3\,$Mpc down to scales somewhat smaller than what is currently accessible by CMB observations \cite{Reid:2009xm}.  Weak gravitational lensing gives access to scales down to $1\,$Mpc \cite{Tegmark:2002cy} while
measurements of the intergalactic hydrogen clumping, traced by the Lyman-$\alpha$ forest,  constrain the density perturbations on scales as small as $\mathcal{O}(10^{-1})\,$Mpc \cite{Tegmark:2002cy,McDonald:2004eu}. Observational evidence for the presence of cosmological perturbations at  smaller scales is essentially absent; the best current limits come from the non-detection of PBHs \cite{GreenLiddle,JGM09}.
Given their sensitivity to the amplitude of small-scale perturbations, UCMHs are potentially much stronger probes of the primordial spectrum at large $k$ than PBHs \cite{SS09,JG10}.  

The most popular and compelling model for DM is that it consists of weakly-interacting massive particles (WIMPs), thermally produced in the early Universe \cite{JKG,Bergstrom:2000pn,Bertone:2004pz,Bertonebook}. 
In this case, the compactness of UCMHs makes them prime targets for indirect detection of self-annihilating DM; 
the non-observation of corresponding gamma rays, in particular, places constraints on the UCMH mass fraction. The most sensitive searches for gamma rays from DM annihilation have been with Air \v{C}erenkov Telescopes (ACTs) \cite{Veritas, Ripken:2010ja, Magic} and the Large Area Telescope (LAT) \cite{LATcluster, LATdwarf, GeringerSameth:2011iw, LATcosmo, Scott:2009jn, Essig:2010em, Abazajian:2010sq,Vertongen:2011mu,Huang:2011xr} aboard the \textit{Fermi} satellite \cite{Atwood:2009ez}.  Another promising technique is to consider the contribution of DM annihilation within UCMHs to the reionisation of the intergalactic medium at redshifts $z\gtrsim6$ \cite{Zhang:2010cj}.  By comparing the predicted Thomson-scattering optical depth from the present day to the surface of last scattering, $\tau_\mathrm{e}$, with the value observed in the CMB, UCMH fractions leading to very early reionisation (and therefore large $\tau_\mathrm{e}$) can be excluded.

In this article, we provide detailed and updated limits on the primordial spectrum at small scales. In particular, we include the non-observation of UCMHs to date by gamma-ray searches for DM with the \textit{Fermi}-LAT, as well as potential impacts of UCMHs upon reionisation.  We expand on previous works in this direction, most notably Ref.~\cite{JG10}, by 
deriving from first principles the minimum density contrast $\delta_\chi^\mathrm{min}$ required to form a UCMH, discussing in detail the transfer and window functions needed to correctly normalise the primordial power spectrum to the one observed today, and by using an improved treatment of the statistics of \textit{Fermi} non-detection of individual sources.
For comparison, we also compile and incorporate other constraints on the primordial power spectrum  into our final limits, extending from the horizon size today down to scales more than 20 orders of magnitude smaller.

We begin in Sec.~\ref{sec:formation} by describing the formation and structure of UCMHs. In Sec.~\ref{sec:abundance}, we then briefly recap the calculation of their cosmological abundance and provide limits on their  present number density.  We continue in Sec.~\ref{sec:constraints} by translating our limits to new constraints on the primordial power spectrum from UCMHs and discussing existing ones, before concluding with Sec.~\ref{sec:conclusions}.  We give a detailed calculation of $\delta_\chi^\mathrm{min}$ in Appendix \ref{app:dmin}, and of the mass variance of perturbations in Appendix \ref{app:sh}, taking into account the correct window and transfer functions.

\section{Formation and structure of UCMHs}
\label{sec:formation}

As noted in Ref.~\cite{Ricotti:2009bs}, density contrasts as small as $\delta\sim10^{-3}$ during radiation domination already suffice to create over-dense regions that would later collapse into UCMHs (see Appendix \ref{app:dmin} for more details). The \emph{initial} mass in CDM particles contained in such a region, i.e.~the mass at the time the corresponding fluctuation of co-moving size $R$ enters into the horizon, is given by 
\bea
 \label{deltam}
 M_i&\simeq&\left[\frac{4\pi}{3}\rho_\chi H^{-3}\right]_{aH=1/R}\nonumber\\
 &=&\frac{H_0^2}{2G}\Omega_\chi R^3\nonumber\\
 &=&1.30\times10^{11}\left(\frac{\Omega_\chi h^2}{0.112}\right)\left(\frac{R}{\rm Mpc}\right)^3 M_\odot\,,
\eea
where $\rho_\chi$ refers to the actual DM density at the time of horizon entry, and $\Omega_\chi$ is the fraction of the critical density of the Universe in DM today. Note that we define $R$ here as the co-moving \emph{radius} of the collapsing region; the co-moving diameter of the region is $2R$, which is \emph{half} the co-moving wavelength of the corresponding physical density fluctuation.  No factors of $g_{\rm eff}$, the effective number of relativistic degrees of freedom at various times in the early history of the Universe, enter in  the above expression, because only the CDM component collapses and contributes to the initial UCMH mass.

During radiation domination, the UCMH-forming `seed' consists only of CDM, and its mass stays essentially constant.
Around matter-radiation equality, it then begins to grow by infall of both dark and (after decoupling) baryonic matter as
\be
\label{Mz}
M_{\rm UCMH}(z)= \frac{z_{\rm eq}+1}{z+1}M_i\,.
\ee
We conservatively assume that this growth only continues until standard structure formation has progressed sufficiently far as to allow star formation, at $z\sim10$.  After this time, dynamical friction between DM halos and hierarchical structure formation presumably make further accretion from the smooth cosmological background of DM inefficient (note that the growth rate presented in Eq.~(\ref{Mz}) assumes accretion in a homogeneous and unbound DM background). Taking $\Omega_\chi h^2 = 0.112$  \cite{wmap}, the mass of a UCMH today is then related to the size of the original, slightly over-dense region by
\be
\label{M0}
M^0_{\rm UCMH}\equiv M_{\rm UCMH}(z\lesssim 10)\approx4\times10^{13}\left(\frac{R}{\rm Mpc}\right)^3 M_\odot\,.
\ee
Similarly, the current-day UCMH mass is related to the horizon mass $M_\mathrm{H}(R)$ at the time when the fluctuation of co-moving size $R$ enters the horizon, by\footnote{
Eq.~(\ref{mhor}) is rigorously correct only when $R\ll (a_\mathrm{eq}H_\mathrm{eq})^{-1}\simeq1.0\times10^2\,{\rm Mpc}$, i.e.~$M_H\ll M_H^{\rm eq}\simeq3.5\times10^{17}M_\odot$; close to $M_H=M_H^{\rm eq}$ it receives a further correction of $2^{-3/4}$.  We give this expression here only for explanatory purposes, and use Eq.~(\ref{deltam}), which is exact, for all the calculations that follow.}
\be
 \label{mhor}
 M^0_{\rm UCMH}\approx3\times10^{-7}\left(\frac{\Omega_\chi h^2}{0.112}\right)
\left(\frac{g_{\rm eff}}{g_{\rm eff}^{\rm eq}}\right)^\frac14
\left(\frac{M_\mathrm{H}}{M_\odot}\right)^\frac32{M_\odot}\,,
\ee
where $g_{\rm eff}$ ($g_{\rm eff}^{\rm eq}$) is evaluated at the time of horizon entry (equality).

After kinetic decoupling \cite{Bringmann:2009vf}, the velocity of CDM particles decreases as $v\propto(1+z)$;
the DM within the over-dense region therefore initially has an extremely low velocity dispersion as long as the density fluctuations are so small that they do not induce sizable gravitational potentials. Even shortly after the onset of matter-domination, when the fluctuations that we are interested in here start to become non-linear, the velocity dispersion increases only mildly as 
\cite{Ricotti:2009bs,Ricotti:2007au}
\be
\label{sigma_v}
\sigma_\mathrm{v}(z)\approx\sigma_{\mathrm{v},0}\left(\frac{1000}{z+1}\right)^\frac12\left(\frac{M_\mathrm{UCMH}(z)}{M_\odot}\right)^{0.28},
\ee
with $\sigma_{\mathrm{v},0}=0.14$\,m\,s$^{-1}$.
UCMHs thus form by almost pure radial infall \cite{Ricotti:2009bs}, which leads to the growth rate presented in Eq.~(\ref{Mz}), and a DM radial profile
\be
\label{UCMHprofile}
\rho_\chi(r,z)=\frac{3f_\chi M_\mathrm{UCMH}(z)}{16\pi R_\mathrm{UCMH}(z)^\frac34r^\frac94},
\ee
where $f_\chi \equiv \Omega_\chi/\Omega_\mathrm{m}$ denotes the fraction of matter that is CDM.
This extremely steep profile, a direct consequence of spherically-symmetric collapse, was first derived analytically \cite{Fillmore:1984wk,Bertschinger:1985pd} and later confirmed by explicit $N$-body simulations \cite{Vogelsberger:2009bn,Ludlow:2010sy} (see in particular Fig.~6 of Ref.~\cite{Vogelsberger:2009bn}).  The factor $f_\chi$ enters because even though the initial UCMH seed consists only of DM, the matter it accretes following decoupling includes both dark and baryonic matter, and accounts for the majority of the mass of the UCMH at the present time.  $R_\mathrm{UCMH}(z)$ refers to the effective radius of the UCMH at redshift $z$, beyond which the density contrast associated with the (fully collapsed) UCMH is $\delta<2$.  For collisionless DM, this turns out to be \cite{Bertschinger:1985pd} 
\begin{equation}
\label{rconversion}
R_\mathrm{UCMH}(z) \approx 0.339 R_\mathrm{ta}(z),
\end{equation}
where the turnaround radius $R_\mathrm{ta}(z)$ is the radius within which matter contained in a collapsing perturbation separates from the Hubble flow.
$R_\mathrm{ta}(z)$ has been obtained by Ricotti \cite{Ricotti:2007jk} from fits to numerical simulations of matter accretion at early times, and then converted by Ricotti, Ostriker \& Mack \cite{Ricotti:2007au} to $R_\mathrm{UCMH}(z)$ using Eq.~(\ref{rconversion}), giving 
\be
\label{UCMHradius}
\frac{R_\mathrm{UCMH}(z)}{\mathrm{pc}}=0.019\left(\frac{1000}{z+1}\right)\left(\frac{M_\mathrm{UCMH}(z)}{M_\odot}\right)^\frac13.
\ee

With our assumed cutoff in accretion at the beginning of star formation at $z\sim10$, the current profiles and radii are obtained by choosing $z\sim10$ in Eqs.~(\ref{UCMHprofile},\ref{UCMHradius}).  Choosing e.g.~$z\sim30$ instead has a minimal impact on results \cite{SS09}.

The steep density profile presented in Eq.~(\ref{UCMHprofile}) is valid so long as the infalling dark matter follows an approximately radial path. This approximation breaks down in the innermost parts of the minihalo, where the average tangential velocity $v_\mathrm{rot}$ of infalling material exceeds the local Keplerian velocity $v_\mathrm{Kep}$.  This occurs only in the inner region because $v_\mathrm{rot}$ rises more steeply with decreasing radius than $v_\mathrm{Kep}$.  The radial infall approximation is hence violated at steadily larger radii as time goes on, as the velocity dispersion of infalling matter increases with time, so matter accreted at later times begins its infall with larger $v_\mathrm{rot}$.  This tends to suppress the contribution of newly-accreted matter to the inner parts of the halo.  Accretion following UCMH formation therefore contributes preferentially to the outer parts of the halo \cite{Bertschinger:1985pd,Ricotti:2009bs}, but leaves the steep inner profile (established by radial infall during the earliest stages of formation) essentially intact.

Even for the earliest-accreted material however, radial infall cannot be valid all the way to $r=0$.  This means that the DM profile in Eq.~(\ref{UCMHprofile}) can only be expected to be valid down to some cutoff radius $r_\mathrm{min}$.  To estimate $r_\mathrm{min}$, we calculate the Keplerian and average tangential velocities of the earliest-accreted material in the UCMH as a function of the radial distance $r$, and then simply solve for the value of $r$ at which these velocities are equal.  With $M(r)$ the total mass contained within radius $r$, from Eq.~(\ref{UCMHprofile}) we have
\be
\label{vkep}
v_\mathrm{Kep}(r) = \sqrt{\frac{GM(r)}{r}} = \sqrt{\frac{{GM_\mathrm{UCMH}(z)}}{r^\frac14R_\mathrm{UCMH}^{3/4}(z)}},
\ee
which we see, from Eqs.~(\ref{Mz}) and (\ref{UCMHradius}), is independent of redshift.  By angular momentum conservation, the mean tangential velocity of the infalling gas is 
\be
v_\mathrm{rot}(r,z) = \frac{\sigma_\mathrm{v}(z) R_\mathrm{UCMH}(z)}{r}\,.
\ee
Setting $v_\mathrm{Kep}$ and $v_\mathrm{rot}$ equal and using Eqs.~(\ref{sigma_v},\ref{UCMHradius}), we find 
\bea
\label{rmin_prelim}
\frac{r_\mathrm{min}}{\rm pc} &\approx&
 4.5\times10^{-5}\left(\frac{1000}{z+1}\right)^{1.33}\left(\frac{R_\mathrm{UCMH}(z)}{\mathrm{pc}}\right)^{0.82}\nonumber\\
&\approx&
 5.1\times10^{-7}\left(\frac{1000}{z+1}\right)^{2.43}\left(\frac{M^0_\mathrm{UCMH}}{M_\odot}\right)^{0.27}\,.
\eea
This gives
\be
\label{rmin}
\frac{r_\mathrm{min}}{R^0_\mathrm{UCMH}} \approx 2.9\times10^{-7}\left(\frac{1000}{z_c+1}\right)^{2.43}\left(\frac{M^0_\mathrm{UCMH}}{M_\odot}\right)^{-0.06}\,,
\ee
where ${R^0_\mathrm{UCMH}}$ is the present-day UCMH radius.
The appropriate redshift to choose here is that at which the UCMH in question collapses, $z=z_c$.  This sets the initial value of the tangential velocity of material collapsing from a height $R_\mathrm{UCMH}(z_c)$ equal to the typical DM velocity $\sigma_\mathrm{v}(z_c)$ at the time of collapse.  As $R_\mathrm{UCMH}(z_c)$ is the maximum height from which matter falls into the UCMH during the initial collapse, this formalism leads to a conservative estimate of $r_\mathrm{min}$.  For a maximally conservative estimate of $r_\mathrm{min}$, one then invokes the largest possible value of $\sigma_\mathrm{v}(z_c)$ at collapse by choosing the smallest allowed redshift of collapse; in our case, this  is $z_c=1000$.

For $r<r_\mathrm{min}$, clearly, the central cusp must be flattened to some extent, as the $r^{-\frac94}$ profile created by radial infall must soften as that approximation breaks down.  The most conservative estimate, which we will follow here, is to simply truncate the profile within $r_\mathrm{min}$, resulting in a cored profile with central density $\rho = \rho_\chi(r_\mathrm{min})$.  Softening the central density profile also has the impact of modifying Eq.~(\ref{vkep}) and hence Eqs.~(\ref{rmin_prelim}) and (\ref{rmin}).  In the most pessimistic case, where the density profile is simply truncated inwards of $r_\mathrm{min}$, the right-hand side of Eqs.~(\ref{rmin_prelim}, \ref{rmin}) should be multiplied by $2^\frac87=2.21$.  In the limit $r_\mathrm{min} \ll R^0_\mathrm{UCMH}$, this results in a reduction of the total gamma-ray flux by a factor of 3.28, and a corresponding increase in the inferred maximum allowed number of UCMHs (see Sec.~\ref{sec:constraints}) by about a factor of 6.  When translated to limits on the primordial power spectrum, such a factor becomes negligible.

The expression in Eq.~(\ref{sigma_v}) for the mean DM velocity dispersion is based on a power-law fit to numerical calculations of $\sigma_\mathrm{v}$ as a function of redshift, within spheres of different co-moving radii \cite{Ricotti:2007au}.  For each redshift and co-moving radius $r_0$, the  velocity dispersion is obtained by performing the integral
\be
 \sigma_\mathrm{v}(z,r_0)^2 = \frac{H(z)^2}{2\pi^2}\int_0^\infty [1-W^2_{\rm TH}(kr_0)]\mathcal{P}_{\delta_\chi}(k,z)\frac{dk}{k^3}\,,
\ee
where $\mathcal{P}_{\delta_\chi}(k,z)$ is the power spectrum of DM density perturbations.  This was computed by the authors of Ref.~\cite{Ricotti:2007au} assuming a Harrison-Zel'dovich spectrum and a standard $\Lambda$CDM cosmology, using earlier code from Ref.~\cite{Bertschinger:1985pd} to track the coupled evolution of the dark and baryonic perturbations.\footnote{
$\mathcal{P}_{\delta_\chi}$ is defined in the analogous way to the total matter power spectrum $\mathcal{P}_{\delta}$, see Eq.~(\ref{pdelta}), but with the replacement of the total matter perturbation $\delta$ by the DM perturbation $\delta_\chi$.  Similarly, it can be obtained in terms of the spectrum of curvature perturbations $\mathcal{P}_\mathcal{R}$ simply by replacing in Eq.~(\ref{pdelta_pr}) 
the radiation transfer function $T_\mathrm{r}$, Eq.~(\ref{Tr}), with the DM transfer function $T_\chi$, Eq.~(\ref{Tx}).} 
$W_{\rm TH}$ is the Fourier-transformed top-hat window function, described in more detail following Eq.~(\ref{pdelta}).  In order to obtain Eq.~(\ref{sigma_v}), the fit is evaluated at a co-moving radius equal to $R_\mathrm{UCMH}(z)$, c.f.~Eq.~(\ref{UCMHradius}).  The fits have a claimed accuracy of 5\%, so the impact of the uncertainty in Eq.~(\ref{sigma_v}) upon our final power spectrum upper limits is negligible for power spectra similar to the Harrison-Zel'dovich case.  In principle however, the DM velocity dispersion could deviate from the behaviour of Eq.~(\ref{sigma_v}) if the calculations of \cite{Ricotti:2007au} were repeated with an alternative input primordial spectrum.  Investigating this effect is beyond the scope of this paper, but might make for interesting future study.

Finally, one other consideration must be taken into account when calculating DM densities in the inner region of UCMHs: over time, WIMP DM will annihilate away, softening the central density cusp.  Following Ref.~\cite{Berezinsky:1992mx} and earlier UCMH work \cite{SS09,JG10}, we estimate $\rho_{_\chi,\mathrm{max}}$, the maximum possible remaining density at time $t$ in a halo born at $t_\mathrm{i}$, as
\be
\label{rcut}
\rho_\chi(r\leq r_\mathrm{cut}) \equiv \rho_{_\chi,\mathrm{max}} = \frac{m_\chi}{\langle \sigma v \rangle (t - t_\mathrm{i})}.
\ee
Here $m_\chi$ is the mass of the WIMP and $\langle \sigma v \rangle$ is its velocity-weighted, thermally-averaged annihilation cross section, taken in the zero-velocity limit.  
For UCMHs seen today, $t=13.76$\,Gyr \cite{wmap};  ignoring gravitational contraction of the DM due to baryonic collapse (seen to have minimal impact upon overall gamma-ray fluxes from UCMHs \cite{SS09}), $t_\mathrm{i} = t(z_\mathrm{eq}) = 59$\,Myr \cite{Wright06}.
We find that $r_\mathrm{cut}\gtrsim4r_\mathrm{min}$ for all the UCMHs we consider in this paper, so their central densities do indeed violate the annihilation bound, even taking into account departures from radial infall.\footnote{The ratio $r_\mathrm{cut}/r_\mathrm{min}$ reduces as the UCMH mass decreases, so in UCMHs lighter than the smallest we consider here ($\sim$$10^{-9}\,M_\odot$) the two radii are comparable.  The crossing point $r_\mathrm{cut}=r_\mathrm{min}$ occurs at $M^0_\mathrm{UCMH}\sim10^{-19}\,M_\odot$ for the parameters we choose here; higher annihilation cross sections or lower DM masses will cause this point to shift to higher $M^0_\mathrm{UCMH}$.}  To account for this, we simply truncate our density profiles instead at $r=r_\mathrm{cut}$, setting the density within this radius equal to $\rho_{_\chi,\mathrm{max}}$. 

Whilst both are important for total annihilation rates, neither the annihilation cutoff nor the correction for violation of radial infall have any real bearing upon the integrated mass of UCMHs.  Correcting e.g.~Eq.~(\ref{UCMHradius}) to retain exactly the same integrated UCMH mass after complete truncation of the density profile inwards of $r_\mathrm{min}$ or $r_\mathrm{cut}$ would result in changes at less than the percent level.

\section{The UCMH population today}
\label{sec:abundance}

\subsection{Present abundance}

For perturbations following a Gaussian distribution, the probability that a region of co-moving size $R$ (at the time when this scale enters the horizon, i.e.~$1/R=aH$) will later collapse into a UCMH is given by
\be
 \label{betafull}
 \beta(R)=\frac{1}{\sqrt{2\pi}\sigma_{\chi,\mathrm{H}}(R)}\int_{\delta_\chi^\mathrm{min}}^{\delta_\chi^\mathrm{max}}\exp\left[-\frac{\delta_\chi^2}{2\sigma_{\chi,\mathrm{H}}^2(R)}\right]\,{\rm d}\delta_\chi\,.
\ee
Here, $\sigma_{\chi,\mathrm{H}}^2$ is the CDM mass variance at horizon entry and $\delta_{\chi}$ is the density contrast in the CDM component only; in Appendix \ref{app:sh}, we provide a detailed recipe for computing $\sigma_{\chi,\mathrm{H}}^2$.  The minimal $\delta_{\chi}$ required for the collapse eventually to happen is given by $\delta_\chi^\mathrm{min}$ (see Appendix \ref{app:dmin} for a precise determination of this value as a function of $R$) and an initial density contrast higher than $\delta_\chi^\mathrm{max}$ would lead to the formation of a PBH rather than a UCMH. PBH formation is expected above $\delta_\chi^\mathrm{max}\sim1/4$ from semi-analytical arguments \cite{pbh_an} or slightly higher values, $\delta_\chi^\mathrm{max}\sim0.5$, from numerical simulations \cite{pbh_num} (recall that for super-horizon adiabatic fluctuations we have $\delta_\chi\sim(3/4) \delta$ during radiation domination). UCMH formation, on the other hand, only requires much smaller density contrasts of the order of $\delta_\chi^\mathrm{min}\sim10^{-3}$ \cite{Ricotti:2009bs}. With $\sigma_{\chi,\mathrm{H}}\sim10^{-5}$, as observed on large scales, we would thus always have $\sigma_{\chi,\mathrm{H}}\ll\delta_\chi^\mathrm{min}\ll\delta_\chi^\mathrm{max}$ and thus
\be
 \beta(R)\simeq\frac{\sigma_{\chi,\mathrm{H}}(R)}{\sqrt{2\pi}\delta_\chi^{\rm min}}\exp\left[-\frac{\delta^{\rm min\,2}_\chi}{2\sigma_{\chi,\mathrm{H}}^2(R)}\right]\,.
\ee
This turns out to be a very good approximation to Eq.~(\ref{betafull}) even in all cases that we will be interested in here, where the power on small scales is significantly larger than on large scales.

Note that $\beta(R)$ also counts those regions of size $R$ that are contained within a larger region $R'>R$ that satisfies $\delta>\delta_\chi^{\rm min}$, too. On the other hand, it does not take into account the possibility that we have $\delta<\delta_\chi^{\rm min}$ for the smaller region $R$, but still have $\delta>\delta_\chi^{\rm min}$ for the larger region $R'$ -- in which case the original region of course would also collapse eventually and end up in a (bigger) UCMH. In the following, we will conservatively neglect these contributions to the total UCMH abundance; in passing, however, we note that in the Press-Schechter formalism \cite{ps} these effects would (somewhat arbitrarily) be accounted for by multiplying the above expressions for $\beta$ by a factor of 2.

Taking into account the accretion of mass described by Eq.~(\ref{Mz}), the present density of UCMHs with mass equal to or greater than $M^0_{\rm UCMH}$ is therefore given by  
\be
 \label{OmegaUCMH}
 \Omega_{\rm UCMH}(M^0_{\rm UCMH})=\Omega_\chi\frac{M^0_{\rm UCMH}}{M_i}\beta(R)\,,
\ee
where $R=R(M^0_{\rm UCMH})$ follows from Eq.~(\ref{M0}).  
Note that this expression does not take into account the potential destruction of UCMHs due to tidal forces and mergers during structure and galaxy formation. 
Similar to the case of super-dense clumps that already collapsed during radiation domination \cite{Berezinsky:2010kq}, however, these effects turn out to be completely negligible.  This is because UCMHs form so early that by the time of structure formation, they have collapsed into quite extreme over-densities with respect to the smooth background.  A good indicator of survival probability is the size of the core radius \cite{Berezinsky:2005py,Berezinsky:2007qu}, given by Eqs.~(\ref{rmin}) and (\ref{rcut}).  A smaller core radius indicates a higher survival probability.  In particular, a core-to-outer radius ratio of less than $\sim$10$^{-3}$ \cite{Berezinsky:2007qu} indicates a survival probability very close to unity; for all the UCMHs we consider in this paper, this ratio is less than $10^{-5}$.  It is also worth recalling that UCMHs evolve as completely isolated objects for some time after they have collapsed: the limits that we will place correspond to rather rare fluctuations with $\delta/\sigma_{\chi,H}\sim3-6$ (relative to a perturbation spectrum where $\sigma_{\chi,H}$ is already enhanced by roughly one order of magnitude on the scale of interest, compared to what is expected from observations at large scales).

We point out that the DM annihilation signal from UCMHs is almost exclusively sensitive to the density in the innermost region; even if UCMHs were to lose part of their outer material due to tidal stripping, this would therefore not affect the corresponding limits that we derive in Section \ref{sec:constraints}. In fact, even for ordinary  DM clumps -- formed in the presence of a standard Harrison-Zel'dovich spectrum of density fluctuations and thus with much smaller densities than UCMHs -- a dense inner core should remain more or less intact and the impact of clump destruction on indirect detection prospects could be much less severe than one naively might expect \cite{Berezinsky:2007qu,Schneider:2010jr}; the impact on UCMHs should be even less. For the following discussion, we thus assume that Eq.~(\ref{M0}) indeed provides a very good estimate for the present UCMH mass, and Eq.~(\ref{OmegaUCMH}) accurately represents the present UCMH density.

As another important consequence of the extremely high density of UCMHs discussed above, we note that the spatial distribution of UCMHs is expected to track the bulk DM. This is quite different from ordinary DM subhalos, which are subject to tidal disruption and therefore generally much less abundant in the centres than outer parts of large halos, relative to the smooth DM component (see e.g.~Ref.~\cite{Springel:2008cc} for a detailed discussion and references). Similarly, we expect the effects of the stellar disk of our Galaxy on the UCMH distribution to be negligible (again in contrast to its effect on ordinary DM clumps, which can be sizable \cite{D'Onghia:2009pz}).

\subsection{Limits from gamma rays: individual Galactic sources}

If UCMHs consist of WIMP DM, they are generically expected to be sources of high-energy gamma rays\footnote{
Note that this is true even in the somewhat contrived situation where WIMPs annihilate, at tree level, only into neutrinos \cite{Ciafaloni:2010ti}.}
(\cite{SS09}; see e.g.~Ref.~\cite{revs} for an overview of gamma-ray yields from WIMP annihilation).  In this case, there exists a unique distance $d_\mathrm{obs}$ out to which UCMHs of any given mass will be observable by gamma-ray telescopes, given a certain instrumental sensitivity.  Using the all-sky gamma-ray survey performed by the \emph{Fermi}-LAT, the present abundance of UCMHs in our own Galaxy can be constrained, based on the non-observation of unassociated point and extended sources with a spectral signature resembling DM annihilation.

Indeed, no unassociated point or extended sources showing evidence of DM annihilation have yet been discovered by \emph{Fermi} \cite{PingApJ,HooperBuckley, Sandick:2010yd, Sandick11}; this is true both for sources in \cite{PingApJ,HooperBuckley} and outside \cite{PingApJ} the 1-year LAT catalogue \cite{1FGL}.  Although Buckley \& Hooper \cite{HooperBuckley} place a rough upper limit of 20--60 on the number of DM halos observed in the 1 year catalogue, many of these can in fact be associated with astrophysical sources; Zechlin et al. \cite{hannes} found 12 possibilities, and then identified the most promising as probably a blazar.  A recent search in the 2-year catalogue \cite{2FGLsources} found just 9 potential sources.  Given that we do not expect all these 9 sources to have been detected at better than 5$\sigma$ in 1 year of data, and that statistically, we expect at least $\sim$80\% of unidentified \emph{Fermi} sources to be relatively easily matched with known sources \cite{Mirabal}, the implied maximum number of UCMHs in the 1-year data is of the order of one or two.  Whilst yet to provide a statistical upper limit on the number of DM halos, the \textit{Fermi}-LAT Collaboration itself reports having seen exactly zero \cite{PingApJ,PingTeVPA,BloomSymp}.  We thus work under the assumption that \emph{Fermi} observed exactly zero UCMHs during its first year of operation, to within its instrumental sensitivity.  Whilst a more detailed treatment would actually include a full spectral analysis of all unassociated sources in the \emph{Fermi} survey, such a procedure is well beyond the scope of this paper (for a full multi-wavelength approach, see Ref.~\cite{hannes}).

The LAT sensitivity to point sources after 1 year of observations, based on a spectral integration above 100\,MeV, is $4\times10^{-9}\,$\,photons\,cm$^{-2}$\,s$^{-1}$ \cite{STwebpage} for a $5\sigma$ detection. Although this sensitivity is based on a power-law spectral source with index $-2$, expected DM annihilation spectra are often sufficiently similar to this that the sensitivity should be broadly similar.  Going beyond this approximation would also require spectral analysis beyond this paper's scope.  We note, however, that pronounced spectral features at high energies, close to the DM particle's mass (in particular from internal bremsstrahlung \cite{Bringmann:2007nk}), can in principle enhance the effective sensitivity by up to an order of magnitude \cite{Bergstrom:2010gh,Bringmann:2011ye}.

In order to derive limits upon the fraction $f$ of Galactic DM contained in UCMHs, let us for simplicity assume that all UCMHs have the same mass $M^0_\mathrm{UCMH}$ -- an assumption we will later comment on. We now pick one particular UCMH in the Milky Way, residing some distance $d$ from Earth.  Assuming that UCMHs track the bulk DM, the probability that this UCMH can be found within a distance $d_\mathrm{obs}$ of Earth is
\begin{equation}
\label{P_1}
P_{d<d_\mathrm{obs},1} = \frac{M_{d<d_\mathrm{obs}}}{M_\mathrm{MW}},
\end{equation}
where $M_\mathrm{MW}$ is the total mass of DM in the Milky Way, and $M_{d<d_\mathrm{obs}}\le M_\mathrm{MW}$ is the mass of DM within $d_\mathrm{obs}$ of Earth.  This probability is simply the fraction of the (dark) Milky Way mass available for the UCMH to turn up in by chance.

The probability of there existing $i$ such UCMHs within $d_\mathrm{obs}$ can then be constructed from the binomial probability of there being a single one, as done in e.g.~Ref.~\cite{Bringmann:2009ip} for intermediate mass black holes.  With the total number of UCMHs of mass $M^0_\mathrm{UCMH}$ in the MW denoted by $N_\mathrm{MW}$, we then have
\be
\label{P_i}
P_{d<d_\mathrm{obs},i} =\binom{N_\mathrm{MW}}{i} (P_{d<d_\mathrm{obs},1})^i (1 - P_{d<d_\mathrm{obs},1})^{N_\mathrm{MW}-i}.
\ee
Because we assume that all UCMHs have the same mass, we can write
\be
N_\mathrm{MW} = f\frac{M_\mathrm{MW}}{f_\chi M^0_\mathrm{UCMH}},
\ee
where we use Eq.~(\ref{OmegaUCMH}) to express the local UCMH mass fraction $f$ in the Milky Way as
\be
\label{f_def}
f \equiv \Omega_\mathrm{UCMH}/\Omega_\mathrm{m}  = \beta(R) f_\chi \frac{M^0_\mathrm{UCMH}}{M_i}.
\ee

In general, the probability that the number of UCMHs $i$ present within $d_\mathrm{obs}$ is equal to or greater than some threshold number $j$, i.e.~the probability that $j$ or more UCMHs exist within $d_\mathrm{obs}$, is simply 1 minus the individual probabilities of there being $0,1,2,\ldots,j-2$ or $j-1$ of them,  
\begin{equation}
P_{d<d_\mathrm{obs},i\ge j} = 1 - \sum_{k=0}^{j-1} P_{d<d_\mathrm{obs},k}.
\end{equation}
So in particular, 
the probability of there being one or more UCMHs within $d_\mathrm{obs}$ is
\begin{align}
\label{P_any}
P_{d<d_\mathrm{obs},i\ge 1} & = 1 - P_{d<d_\mathrm{obs},0}\nonumber\\
              & = 1 - \left(1-\frac{M_{d<d_\mathrm{obs}}}{M_\mathrm{MW}}\right)^{\frac{fM_\mathrm{MW}}{f_\chi M^0_\mathrm{UCMH}}}.
\end{align}
This equation provides an estimate of the probability of there being any UCMHs at all within an observable distance, as a function of the UCMH mass, their maximum observable distance, and the UCMH mass fraction.  If we know the sensitivity for a source detection at confidence level (CL) $x$, then there is a probability $x$ that a UCMH residing at exactly $d_\mathrm{obs}$, with a mean flux equal to the sensitivity, would be seen by \emph{Fermi}.  The total probability that we would observe one or more UCMHs with \textit{Fermi} is then $P_\mathrm{tot} = xP_{d<d_\mathrm{obs},i\ge 1}$.  To derive an upper limit on $f$ from a non-observation of UCMHs by \textit{Fermi}, such that $f<f_\mathrm{max}$ at some CL $y$, we therefore require that $P_\mathrm{tot}>y$ for $f>f_\mathrm{max}$.  That is, $y = \left. x P_{d<d_\mathrm{obs},i\ge1}\right|_{f=f\mathrm{max}}$, which gives
\be
\label{fmax}f_\mathrm{max} = \frac{f_\chi M^0_\mathrm{UCMH}}{M_\mathrm{MW}}\frac{\log\left(1-\frac{y}{x}\right)}{\log\left(1-\frac{M_{d<d_\mathrm{obs}}}{M_\mathrm{MW}}\right)}\,.
\ee 
For close objects, $M_{d<d_\mathrm{obs}}\ll M_\mathrm{MW}$, and we have
\be
 \label{fclose}
 f_\mathrm{max}\simeq - \frac{f_\chi M^0_\mathrm{UCMH}}{M_{d<d_\mathrm{obs}}}\ln\left(1-\frac{y}{x}\right)\,,
\ee
which is independent of $M_\mathrm{MW}$, as expected.  This expression has a rather intuitive meaning: it gives the mass fraction a single UCMH would contribute to the total DM mass within $d_{\rm obs}$, times a statistical weighting function, which accounts for the fact that a non-observation of UCMHs does not guarantee that there indeed are none in the observable volume, nor does it guarantee that such a volume would always contain zero such minihalos, even were we certain that this particular volume does not.

In terms of the gamma-ray flux $\mathcal{F}(d)$ from a single UCMH at some distance $d$, this becomes
\be
\label{fmax_fluxes}
 f_\mathrm{max}\simeq-\left[\frac{\mathcal{F}_\mathrm{min}}{\mathcal{F}(d)}\right]^\frac32\frac{3f_\chi M^0_\mathrm{UCMH}}{4\pi\rho_\chi d^3}\log\left(1-\frac{y}{x}\right),
\ee
where $\rho_\chi $ is the local DM density and the point-source sensitivity is defined as $\mathcal{F}_\mathrm{min} \equiv \mathcal{F}(d_{\rm obs})$.  Note that this expression is in fact independent of $d$, as $\mathcal{F}(d)\varpropto d^{-2}$.

For a spherically-symmetric DM halo appearing as a point source at distance $d$, the observed gamma-ray flux integrated above some threshold energy $E_\mathrm{th}$ is 
\begin{equation}
\label{ptsrc}
\mathcal{F}(d) = \sum_k\int^{m_\chi}_{E_\mathrm{th}} \frac{\mathrm{d}N_k}{\mathrm{d}E}\mathrm{d}E \frac{\langle\sigma_k v\rangle}{2d^2m_\chi^2} \int_0^{R^0_\mathrm{UCMH}}r^2\rho^2(r)\mathrm{d}r.
\end{equation}
Here $\langle \sigma_k v \rangle$ is the cross section of the $k$th annihilation channel and $\mathrm{d}N_k/\mathrm{d}E$ is its differential photon yield.  Inverting this expression gives $d_\mathrm{obs}$ as a function of $\mathcal{F}_\mathrm{min}$.

In order to calculate $M_{d<d_\mathrm{obs}}$, we integrate the smooth DM profile of the Milky Way over a sphere of radius $d_\mathrm{obs}$ around the Sun, using the NFW profile of Ref.~\cite{Battaglia06} ($c=18$, $M_{200}=9.4\times10^{11}\,M_\odot$, $r_\mathrm{s}=17.0$\,kpc).  In principle, local DM substructure in the solar vicinity could make this a biased estimator of $M_{d<d_\mathrm{obs}}$ for small $d_\mathrm{obs}$. Current understanding of the granularity of DM halos unfortunately does not allow the magnitude of this effect to be accurately assessed.  Recent N-body simulations \cite{Zemp:2008gw} indicate a $\sim$10\% variation in density over a sphere of 500\,pc around the Sun, and more over somewhat smaller spheres.  These scales are at the limit of the simulation's resolution however, so going to smaller spheres also requires higher resolution, which in turn reduces the observed variation (see the discussion in Sec.~3.2 and the Appendices of Ref.~\cite{Zemp:2008gw}).  It is not obvious which effect would win out at the smallest scales we consider here, which are another 6 orders of magnitude smaller.

We have constructed Eqs.~(\ref{P_1}--\ref{fmax_fluxes}) assuming that \textit{all} UCMHs in the Milky Way have the same mass.  Because we have observed exactly zero UCMHs of any mass however, the resulting limits on $f$ in fact immediately generalise to arbitrary UCMH mass spectra.  This is because we know exactly how many UCMHs of each mass have been observed (none), so the limits for different masses can all be applied independently.  If we were instead using some $j$ observed UCMH candidate sources to draw an upper limit on $f$ (i.e.~rather than using them to claim a UCMH detection), we would have a form of Eq.~(\ref{P_any}) where $i\ge j$ for some $j\ge2$.  In this case, we would have to specify the form of the UCMH mass spectrum in order to know what fraction of observed sources to attribute to each mass band considered in the analysis.  Even in this case though, the most conservative limits would be obtained by assuming, for each band individually, that all the observed sources were UCMHs with masses in that band.

\begin{figure}[t]
\includegraphics[width=\columnwidth, trim = 0 20 0 0, clip=true]{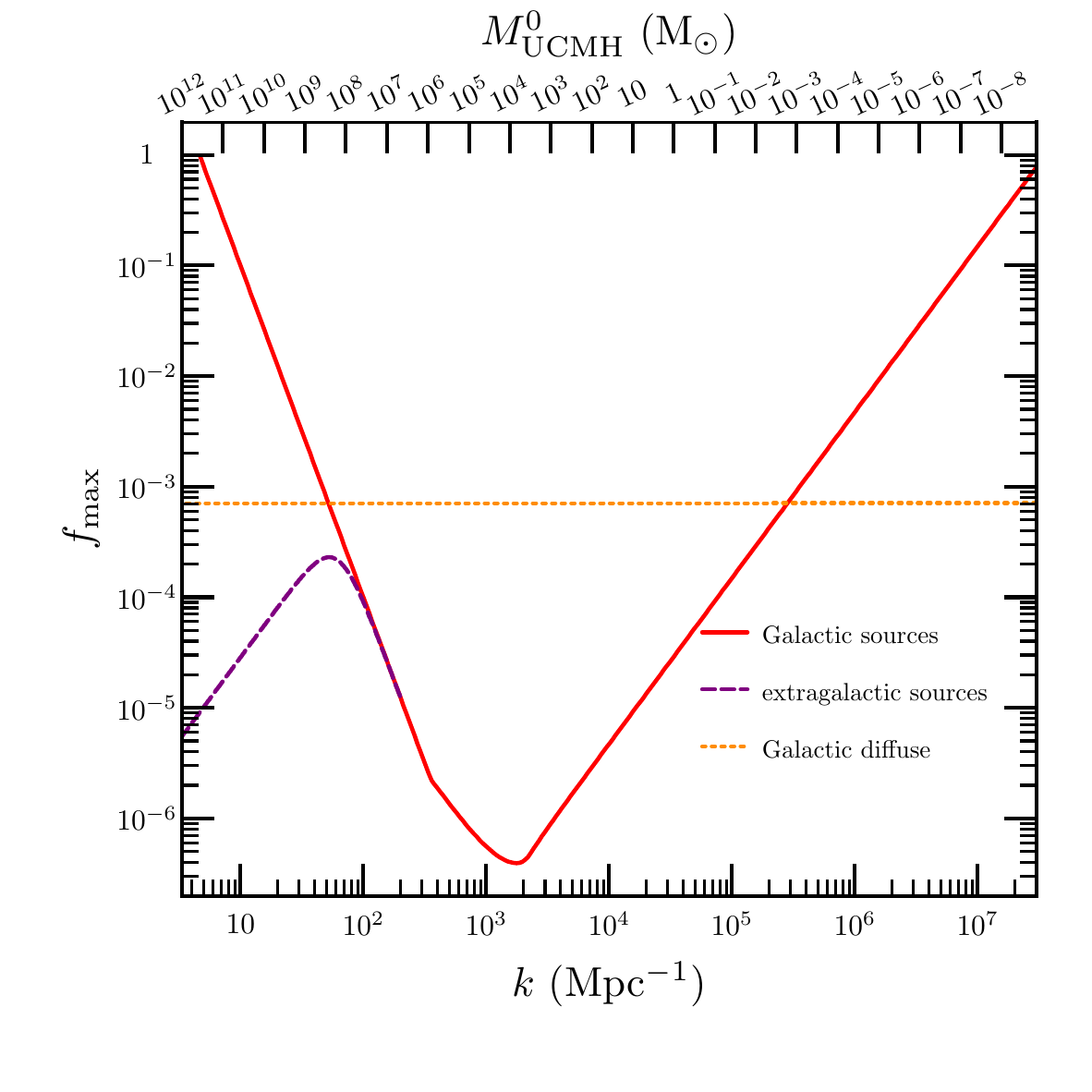}
\caption{The maximum allowed fraction of DM in the Milky Way contained in UCMHs, as a function of $k$ and the UCMH mass $M^0_\mathrm{UCMH}$.  Here we show limits derived in this paper from \textit{Fermi}-LAT searches for individual and diffuse DM sources.  The UCMH mass is related to the mass contained inside the horizon when mode $k$ enters by Eq.~(\ref{mhor}).  All limits correspond to a 95\% CL.  Limits from searches for individual minihalos  are based on non-observation of point or extended DM sources during one year of operation in all-sky survey mode.}
\label{fig:fUCMH_constraint}
\end{figure}

Using Eq.~(\ref{fmax}) and the integrated LAT sensitivity discussed above, we have determined the 1-year, 95\% CL \emph{Fermi}-LAT upper limits on $f$, the fraction of Galactic DM contained in UCMHs of each mass.  This limit is shown in Fig.~\ref{fig:fUCMH_constraint} as a solid red line.  Here we have calculated integrated gamma-ray fluxes above photon energies of 100\,MeV, using a suitably extended version of DarkSUSY \cite{darksusy}\footnote{UCMH routines will be included in a future public release of DarkSUSY.}, as described in \cite{SS09}.

We assume 100\% annihilation of WIMPs into $b\bar{b}$ pairs, a WIMP mass of $m_\chi=1$\,TeV and an effective annihilation cross-section of $\langle\sigma v\rangle=3\times10^{-26}$\,cm$^3$\,s$^{-1}$.  These are fairly conservative choices as far as gamma-ray yields go.  Heavier WIMPs give lower fluxes, and WIMP masses considerably higher than 1\,TeV are extremely challenging to obtain if one hopes for an associated natural solution to the gauge hierarchy problem.  Our annihilation cross-section is the canonical, unboosted $s$-wave value implied by the relic density of DM under the assumption of thermal production.  The $b\bar{b}$ final state gives rise to a relatively soft continuum spectrum dominated by pion decay; significant yields into final states with lower gamma-ray yields, such as $\mu^+\mu^-$, typically only arise in WIMP models engineered to explain cosmic-ray excesses, and are in those cases accompanied by a corresponding boost factor in the annihilation rate.  Taking both these effects into account, integrated UCMH gamma-ray fluxes from models annihilating into $\mu^+\mu^-$ are not enormously different to those arising from unboosted annihilation into $b\bar{b}$ \cite{SS09}. 

Reading from right to left in Fig.~\ref{fig:fUCMH_constraint}, the Galactic gamma-ray source limit strengthens with increasing UCMH mass as UCMHs become brighter and more of the Galaxy is contained within $d_\mathrm{obs}$.  At a mass of $\sim$$7 \times 10^3$\,$M_\odot$, the limit is strongest, and UCMHs must constitute less than about $4\times10^{-7}$ of all DM in the Milky Way.  

At masses above $\sim$$10^6\,M_\odot$, the value of $f_\mathrm{max}$ given by Eq.~(\ref{fmax}) corresponds to less than three UCMHs in the entire Milky Way.  At this point, Eq.~(\ref{fmax}) breaks down due to low-number statistics; the Milky Way may simply contain zero UCMHs of a given mass in this case purely by chance, even though they are cosmologically more abundant than Eq.~(\ref{fmax}) would suggest.  For larger masses, we obtain constraints by assuming that on average there are at most 3 UCMHs of the mass in question per Milky-way sized halo.  For zero observed UCMHs of a given mass, the CL with which we exclude a model that gives on average $n$ such UCMHs per  Milky-way sized halo is $x[1 - \exp(-n)]$, with $x$ the confidence level of the observation as in Eqs.~(\ref{fmax}--\ref{fmax_fluxes}); choosing $n=3$ makes this is a 95\% confidence exclusion.  From the minimum at intermediate UCMH mass, larger masses lead to less stringent limits on $f$, as three UCMHs progressively occupy a larger fraction of the mass of the Milky Way halo.

The limiting behaviour in Fig.~\ref{fig:fUCMH_constraint} at very large and very small masses is simple to understand.  For large UCMH masses, $M^0_\mathrm{UCMH}$ approaches the Milky Way mass, and because the limit here is given by the assumption that no more than three UCMHs of a given mass exist in each Milky-Way size halo, $f_\mathrm{max}$ approaches one at exactly $M^0_\mathrm{UCMH}={M_\mathrm{MW}}/{3f_\chi}$.

For small UCMH masses, $M_{d<d_\mathrm{obs}}$ eventually shrinks to such an extent that Eqs.~(\ref{fmax}) and (\ref{fmax_fluxes}) become greater than one.  Common sense of course dictates that UCMHs cannot make up more than 100\% of the Milky Way mass, but in our formalism this knowledge does not place any limit on the size of cosmological perturbations at such large $k$.  This is because the \emph{Fermi} limit is `saturated' with respect to perturbations of this size; the perturbations can be arbitrarily large and still (by definition) unable to cause over 100\% of the mass of the Milky Way to reside in UCMHs.  For this reason, we do not give any limits for wavenumbers where $f_\mathrm{max}=1$.  For searches for individual Galactic gamma-ray sources, this corresponds to UCMH masses below $\sim$$10^{-7}$\,$M_\odot$.  In principle, one could obtain some bounds for larger wavenumbers by relaxing Eq.~(\ref{Mz}), and deriving direct limits on $\beta$ as a function of $M_i$, in cases where large amplitude density fluctuations would result in $f=1$ before $z=10$.  This would be rather brave however, as it is not known to what extent the radial infall and absolute survival approximations, which we rely on here, should be violated close to $f=1$.

Such masses are already well into the regime in which kinetic coupling of DM might be expected to wash out structures such as UCMHs anyway.  Indeed, this is an important general caveat at low masses: depending upon the specific particle DM candidate, the resultant mass cutoff for UCMHs can be many orders of magnitude larger than the smallest masses we consider here.  We urge the reader to remember that the limits we present do not apply below the WIMP kinetic decoupling threshold, and that this threshold should be calculated on a per-model basis \cite{Bringmann:2009vf}.  For e.g.~neutralino DM, the corresponding size of the smallest DM halos at the time of equality may in principle be anything between $M_i=10^{-11}$\,$M_\odot$ and $M_i=10^{-3}$\,$M_\odot$ \cite{Bringmann:2009vf}, corresponding to $k_{\rm max}^\chi\sim 10^8-10^5\,{\rm Mpc}^{-1}$  (although it is often assumed to be $\sim$$10^{-6}$\,$M_\odot$, as for a 100\,GeV pure Bino \cite{Hofmann:2001bi,Green:2003un,Green:2005fa}).

\subsection{Limits from gamma rays: individual extragalactic sources}

It is straightforward to extend the Galactic source analysis to sources residing outside the Milky Way.  The approximate expression for $f_\mathrm{max}$ close to the Earth, Eq.~(\ref{fclose}), in fact holds in this case, as the approximation relies only on the probed volume being much smaller than that in which the number of UCMHs was assumed to be fixed.\footnote{Formally, one can replace $M_\mathrm{MW}$ in Eq.~(\ref{fmax}) with e.g. the mass of the local group, cluster or supercluster, and make the same Taylor expansion as required to obtain Eq.~(\ref{fclose}).}  In this case, the mass contained within $d_\mathrm{obs}$ is given by 
\be
  M_{d<d_\mathrm{obs}}=\frac{4\pi}{3}\left(d_{\rm obs}^3-d^3_{max, MW}\right)\rho^0_c\Omega_\chi+M_{MW}\,,
\ee
where  $\rho^0_c\Omega_\chi$ is the local cosmological background DM density.
We show the resulting values of $f_\mathrm{max}$ in Fig.~\ref{fig:fUCMH_constraint} as a dashed purple line.  As expected, the extragalactic limits provide increased sensitivity in the large-mass region, where the Galactic limits are set by the $n=3$ condition.  Having constructed both the Galactic and extragalactic limits as 95\% CL upper limits, we see that the extragalactic curve tracks the Galactic one where $d_\mathrm{obs}$ is comparable to $R_\mathrm{MW}$.  At smaller $k$, the extragalactic limit turns over and becomes stronger again, as unlike the Galactic DM profile, the cosmological background density is approximately constant with increasing $d$, providing a volume boost to the sensitivity for increasing UCMH mass.

\subsection{Limits from gamma rays: Galactic diffuse emission}

For very small UCMH masses, where $d_\mathrm{obs}$ is small but UCMHs are potentially very numerous, one might expect  the diffuse gamma-ray background to provide a stronger limit on the UCMH fraction than searches for individual sources.  In this case, a simple and robust limit can be obtained by considering the gamma-ray flux from a direction perpendicular to the Galactic disk, as this is least likely to be contaminated by astrophysical sources like pulsars and supernova remnants.  The flux one would expect from unresolved Galactic UCMHs in this direction, integrated in energy but differential in solid angle $\Omega$, is
\be
  \label{diffspec}
  \frac{\mathrm{d}\mathcal{F_{\rm diff}}}{\mathrm{d}\Omega} = \frac{f}{f_\chi M^0_{\rm UCMH}}\int_0^{d_\mathrm{max}}\rho_\chi(d)\mathcal{F}(d)d^2 {\rm d}d\,, 
\ee
with $d_\mathrm{max} \equiv {\sqrt{R_{\rm MW}^2 + r_0^2}}$. Here, $R_{\rm MW}$ is the virial radius of the Milky Way and $r_0$  the distance of the Sun from the Galactic centre; we take $R_{\rm MW}=305\,$kpc \cite{Battaglia06} and $r_0=8\,$kpc.  Eq.~(\ref{diffspec}) holds whether one refers to the raw flux $\mathcal{F}(d)$ or the energy flux $\mathcal{F}_\mathrm{E}(d)$, which differs from the raw flux only by an additional weighting factor of $E$ inside the energy integral of Eq.~(\ref{ptsrc}).

To obtain conservative upper limits on $f$, we compare this prediction to the total diffuse gamma-ray energy flux observed by \textit{Fermi} at the Galactic poles.  Referring to e.g. Fig.~1 in the 2-year LAT bright source catalogue \cite{2FGL}, this is about $1\times10^{-5}$\,GeV\,cm$^{-2}$\,s$^{-1}$\,sr$^{-1}$.  We derive 95\% CL upper limits on $f$ by demanding that the diffuse energy flux from UCMHs in this direction does not exceed the total measured value times 1.2, which corresponds to an upwards variation of two times $\sigma_\mathrm{eff}=10\%$, the maximum systematic error in the LAT effective area over its energy range \cite{2FGL}.  We implicitly assume in this procedure that statistical variation of the polar flux over the 2-year observing period is below the level of the systematic error coming from the LAT effective area, a reasonable approximation for our purposes.

\begin{figure}[t]
\includegraphics[width=\columnwidth, trim = 0 20 0 0, clip=true]{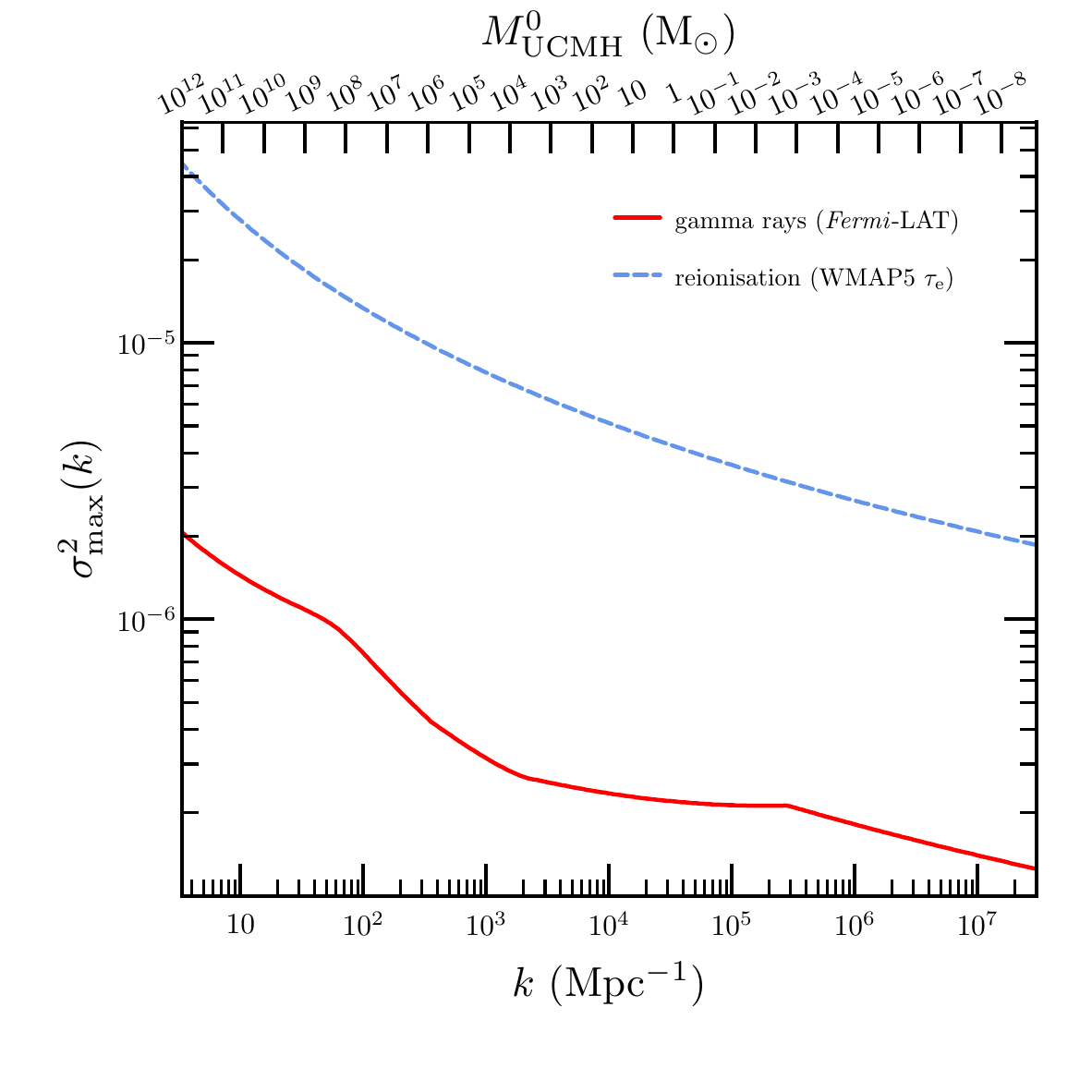}
\caption{Upper limits on the mass variance $\sigma^2_{\chi,\mathrm{H}}$ at horizon entry ($aH=k$), implied by the present-day UCMH abundance limits presented in Fig.~\ref{fig:fUCMH_constraint}, as well as the limits on $f$ at the time of matter-radiation equality derived from the CMB, see Eq.~(\ref{reion}).  This later limit refers to the impact of UCMHs upon reionisation \protect\cite{Zhang:2010cj}; larger values of $\sigma^2_{\chi,\mathrm{H}}$ correspond to a UCMH fraction that speeds up reionisation to the point where the integrated optical depth of the CMB ($\tau_\mathrm{e}$) is not consistent with the value measured by WMAP5 \protect\cite{wmap5}.}
\label{fig:sigH_constraint}
\end{figure}

From the upper limits on $f$ presented in Fig.~\ref{fig:fUCMH_constraint}, we are able to determine upper limits on the mass variance of perturbations using Eqs.~(\ref{betafull}) and (\ref{f_def}), assuming that the fractional UCMH content of the Milky Way is indeed indicative of DM in the rest of the Universe.  We use
Brent's Method\footnote{Refer to standard numerical texts such as \cite{NR} for a description of all algorithms mentioned in this paper.} to numerically determine the value $\sigma^2_{\chi,\mathrm{H}}=\sigma^2_\mathrm{max}$ corresponding to $f_\mathrm{max}$, and show the resultant upper limits on $\sigma^2_{\chi,\mathrm{H}}$ in Fig.~\ref{fig:sigH_constraint}.  For this figure, we have combined the three different \textit{Fermi}-LAT gamma-ray limits just discussed into a single best limit.

\subsection{Limits from reionisation}

We also plot in Fig.~\ref{fig:sigH_constraint} the upper limit on $\sigma^2_{\chi,\mathrm{H}}$ based on the limit on $f$ at the time of matter-radiation equality, calculated by Zhang \cite{Zhang:2010cj}:
\begin{equation}
\label{reion}
f_\mathrm{eq}\lesssim 10^{-2}(m_\chi/100\,\mathrm{GeV}).
\end{equation}
This limit comes from the impact of DM annihilation in UCMHs upon reionisation and the integrated optical depth of the CMB at $z<30$ (as seen by WMAP5 \cite{wmap5}).  For larger $f_\mathrm{eq}$, reionisation takes place at higher $z$, introducing free electrons to the intergalactic medium at earlier times and increasing the Thomson-scattering optical depth to the surface of last scattering ($\tau_\mathrm{e}$).  We implemented this limit with the same DM mass of 1\,TeV discussed above.  According to Eqs.~(\ref{UCMHprofile}), (\ref{UCMHradius}) and (\ref{ptsrc}), the number of ionising photons $N_\gamma$ produced by a single UCMH scales linearly with its mass, as 
\be
\label{nionphot}
N_\gamma\ \varpropto\ \mathcal{F}(d)d^2\ \varpropto\ \int_0^{R^0_\mathrm{UCMH}}r^2\rho^2(r)\mathrm{d}r\ \varpropto\ M^0_\mathrm{UCMH}.
\ee
For a fixed $f$ the number of UCMHs scales inversely with their mass, so we see that the limit in Eq.~(\ref{reion}) should be independent of the UCMH mass.  This limit is weaker than those from gamma rays.  We do not give this limit in Fig.~\ref{fig:fUCMH_constraint}, as such a UCMH fraction at equality would grow to become simply $f=1$ today (i.e. the bound saturates before $z=0$).

\subsection{Limits from gravitational lensing}

Various gravitational microlensing experiments looking towards the Large and Small Magellanic clouds have been carried out in recent times \cite{Alcock:1993eu,Aubourg:1993wb,Udalski:1993zz,Yock1998,Kerins2008,Auriere2001,Riffeser:2003rs}.  The strongest resulting limits to date on the Milky Way halo mass fraction contained in massive compact halo objects (MACHOs)~\cite{Paczynski:1985jf} have been provided by EROS and OGLE (see e.g.~\cite{Wyrzykowski:2010mh,Wyrzykowski:2011tr}), on scales of $k \sim 10^{4}$ Mpc$^{-1}$ to $k \sim 10^{7}$ Mpc$^{-1}$. 

UCMHs are in fact just a non-baryonic variant of MACHOs. Naively applying these limits, which are very roughly of the order of $f_{\rm max}\sim 0.1$, would provide constraints on the primordial spectrum of density fluctuations (see Section \ref{sec:constraints}) that are only slightly weaker than the constraints we derive from the gamma-ray limits. More importantly, these limits would not rely on the WIMP hypothesis and thus be completely independent of the nature of DM.

As it turns out, however, UCMHs cannot be expected to have shown up in the EROS or OGLE data analysis because they are not sufficiently point-like. Indeed, their mass is much larger than the mass contained inside their Einstein radius, which significantly reduces the expected magnification in lensing events. 
For this reason, we do not include any limits from gravitational lensing in our analysis.  We note however that
the light curve expected from UCMH lensing could help to detect such events even against a large background of non-MACHO events  \cite{Ricotti:2009bs}, an effect which may produce interesting limits for future photometric lensing searches with, e.g., \textit{Kepler} \cite{Griest11}. Another very promising way to constrain the abundance of UCMHs is to use astrometric microlensing, which involves searching for small changes in the apparent position of background stars with future planet-hunting facilities \cite{Erickcek10,Erickcek12}.


\section{Constraints on the primordial power spectrum}
\label{sec:constraints}

\begin{figure}[t]
\includegraphics[width=\columnwidth, trim = 0 20 0 0, clip=true]{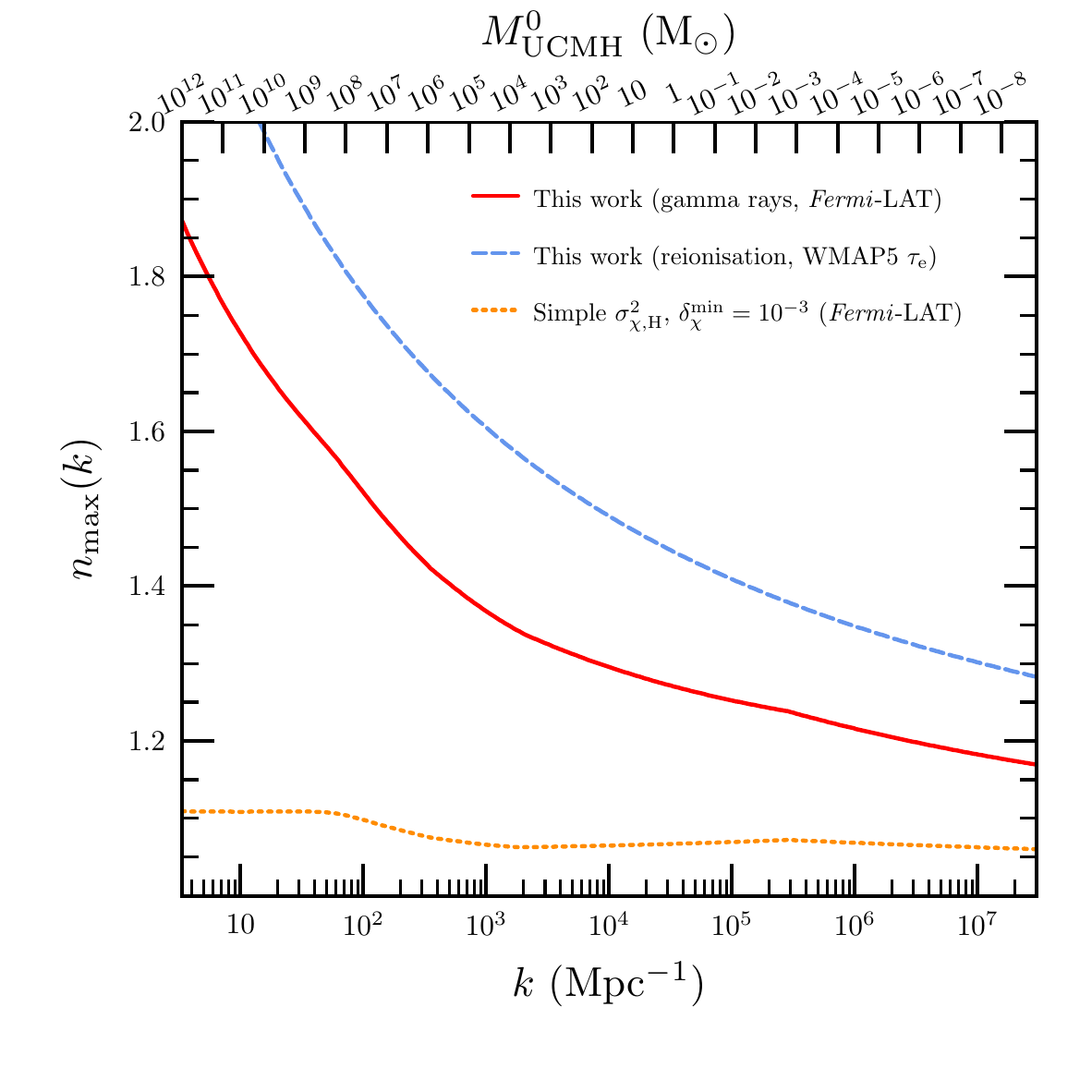}
\caption{Upper limit on the spectral index of the primordial power spectrum from gamma-ray searches for UCMHs, and the impact of UCMHs on reionisation.  These limits assume $\delta^2_\mathrm{H}\propto k^{n-1}$, and take into account only the constraints on $\sigma_{\chi,\mathrm{H}}$ given in Fig.~\ref{fig:sigH_constraint}, for wave numbers smaller than $k$.  For comparison, we also show the resulting gamma-ray constraint if we were to assume $\delta_\chi^\mathrm{min}=10^{-3}$ (improved upon in Appendix \ref{app:dmin}) and use the over-simplified calculation of $\sigma_{\chi,\mathrm{H}}$ \cite{GreenLiddle} (corrected in Appendix \ref{app:sh}).}
\label{fig:n_constraint}
\end{figure}

\begin{figure*}[t]
\includegraphics[width=\columnwidth, trim = 0 20 0 0, clip=true]{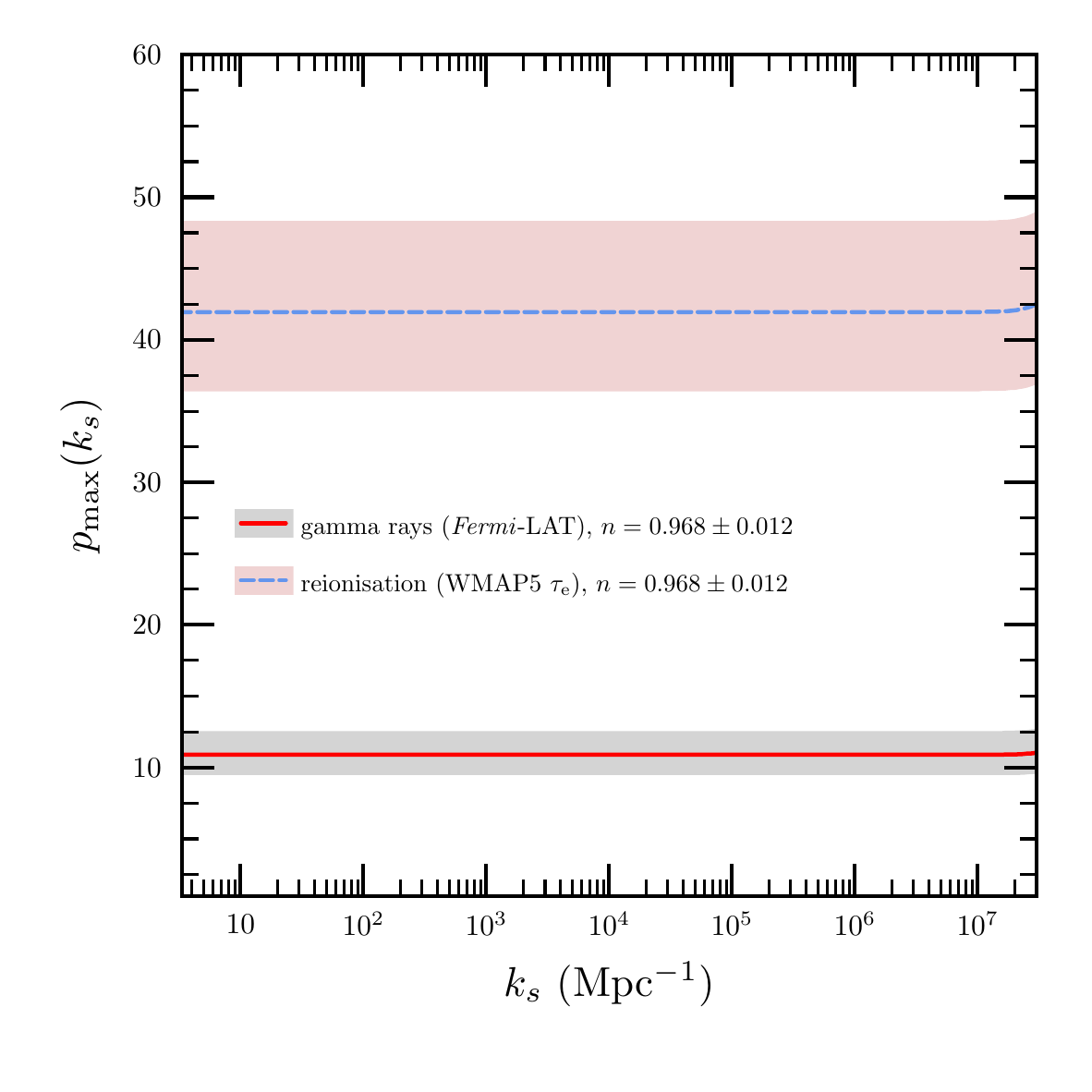}\hspace{5mm}
\includegraphics[width=\columnwidth, trim = 0 20 0 0, clip=true]{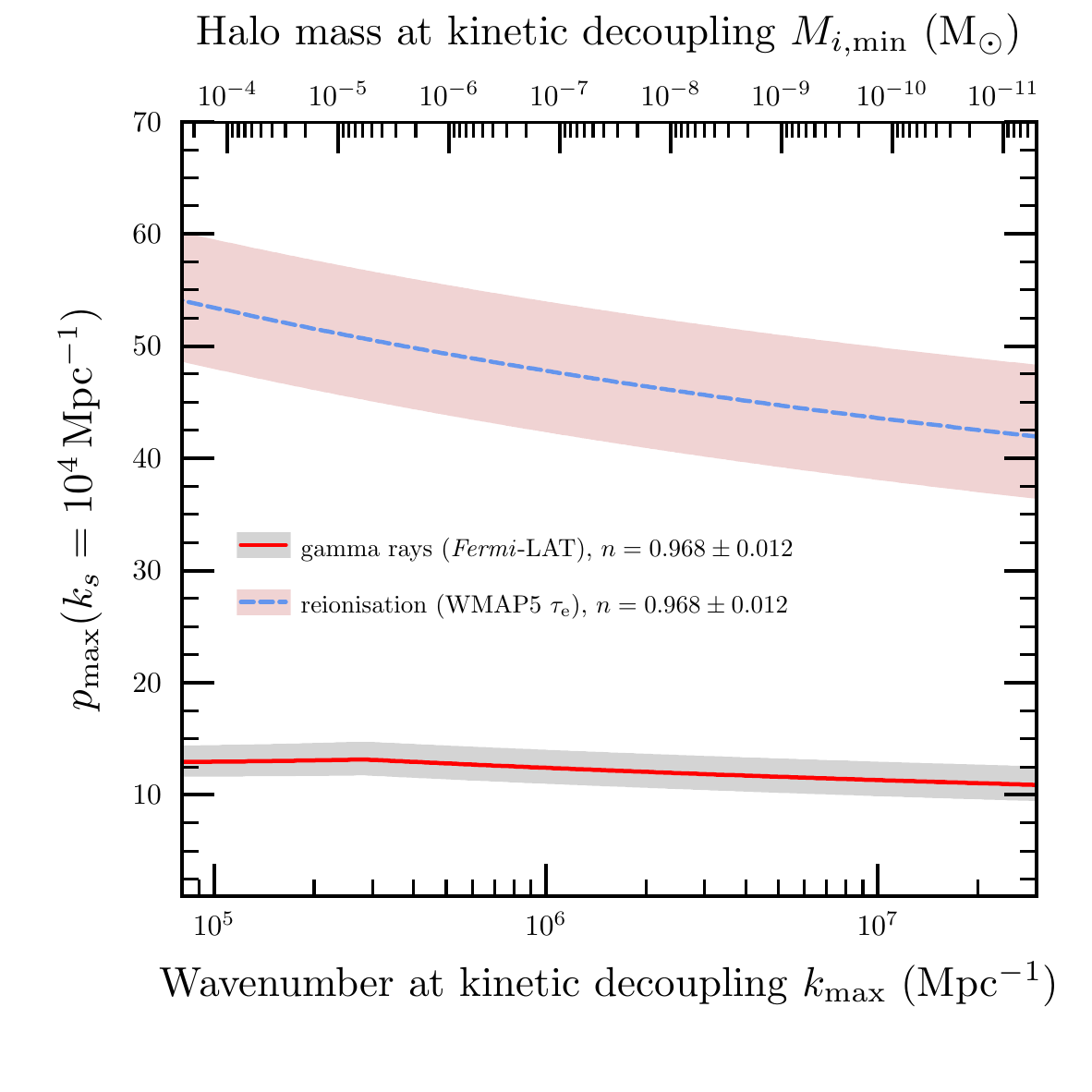}
\caption{\emph{Left:} Constraint on the allowed height $p$ of a step in the primordial power spectrum from gamma-ray searches for UCMHs and impacts on reionisation, as a function of the location $k_s$ of such a step.  Here $p$ refers to the dimensionless ratio of the power at the wavenumbers immediately above and below $k_s$.  Our central curves assume the spectral index $n=0.968$ obtained from WMAP7 observations of large scales \cite{wmap}, and shaded regions correspond to the 68\% CL for this measurement ($\Delta n=0.012$). \emph{Right:} Variation of the gamma-ray and reionisation constraints on $p$ with the kinetic decoupling scale of DM.  These two limits in particular are sensitive to the cutoff in the DM halo mass function, as the strongest limits (as shown in the left panel) come from the smallest viable UCMHs, for all $k_s$.}
\label{fig:p_constraint}
\end{figure*}

\begin{figure*}[t]
\includegraphics[width=\columnwidth, trim = 0 20 0 0, clip=true]{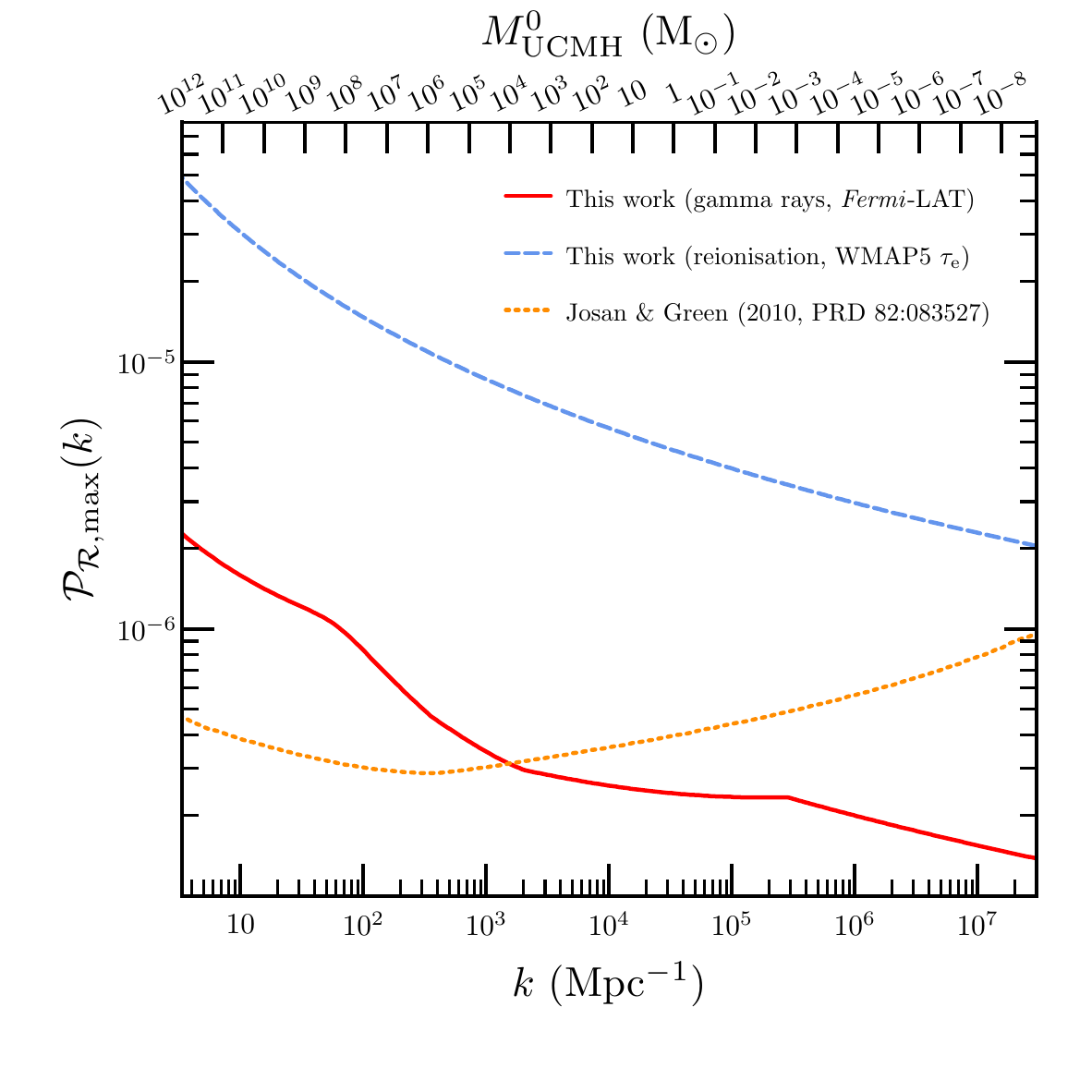}\hspace{5mm}
\includegraphics[width=\columnwidth, trim = 0 20 0 0, clip=true]{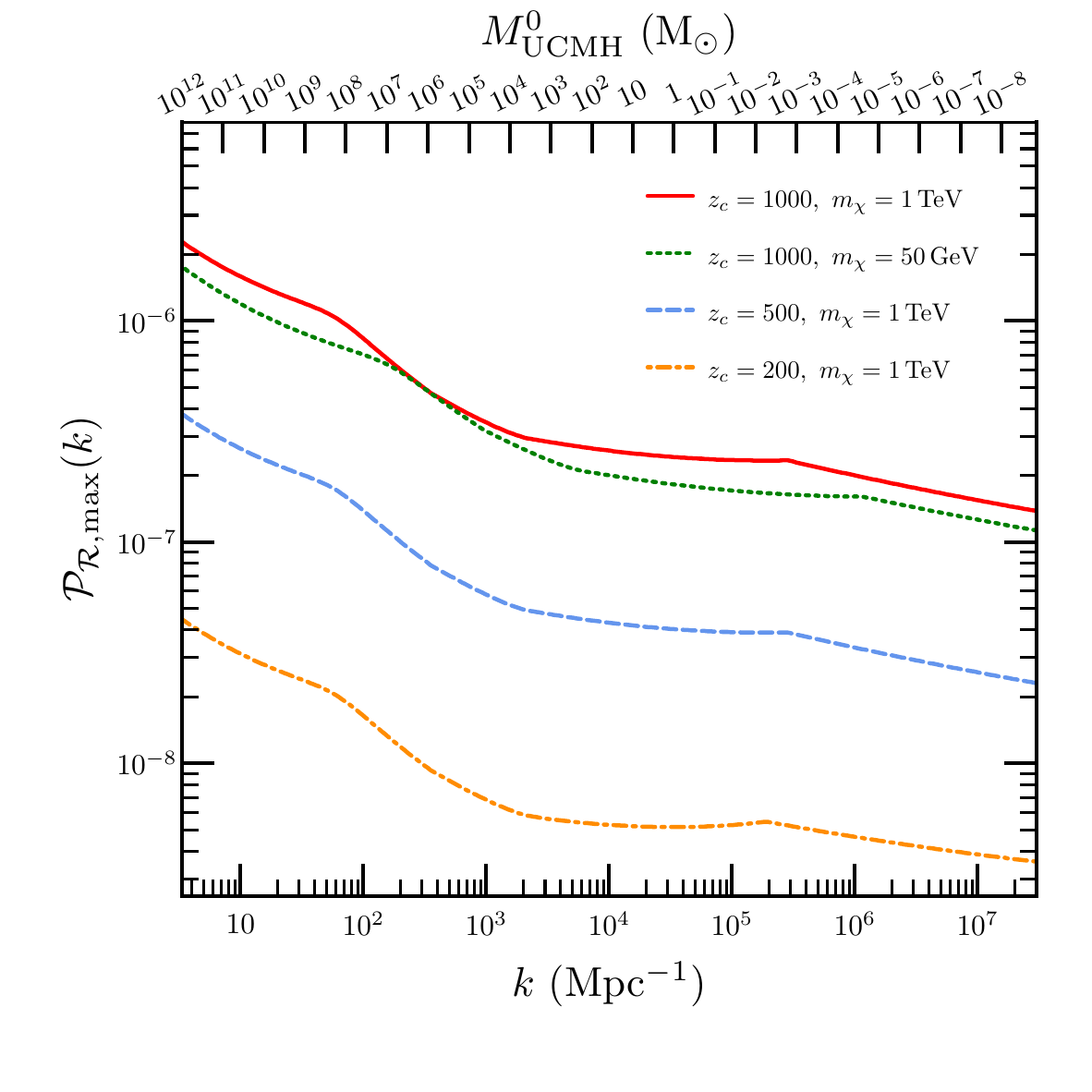}
\caption{\emph{Left:} 95\% CL upper limits on the amplitude of primordial curvature perturbations $\mathcal{P}_\mathcal{R}$ (for a non-parametric, generalised spectrum) allowed by gamma-ray searches for UCMHs and impacts on reionisation.  For comparison, we show the previous limits from \emph{Fermi} non-observation of UCMHs derived in Ref.~\cite{JG10}, based on a simplified treatment of the statistics of non-detection, mass variance and minimum density contrast required to form a UCMH.  Corresponding constraints on the generalised amplitude of primordial density perturbations can be obtained by multiplying these limits by a factor of 0.191, see Eq.~(\ref{pdelta_conversion}). \emph{Right:} The variation of the gamma-ray limit on $\mathcal{P}_\mathcal{R}$ with WIMP mass and the redshift of UCMH collapse, showing the impact of less conservative (but entirely plausible) choices for these parameters than our canonical $m_\chi=1$\,TeV, $z_c=1000$.}
\label{fig:PR_constraint}
\end{figure*}

\begin{figure*}[t]
\includegraphics[width=\linewidth, trim = 0 92 0 100, clip=true]{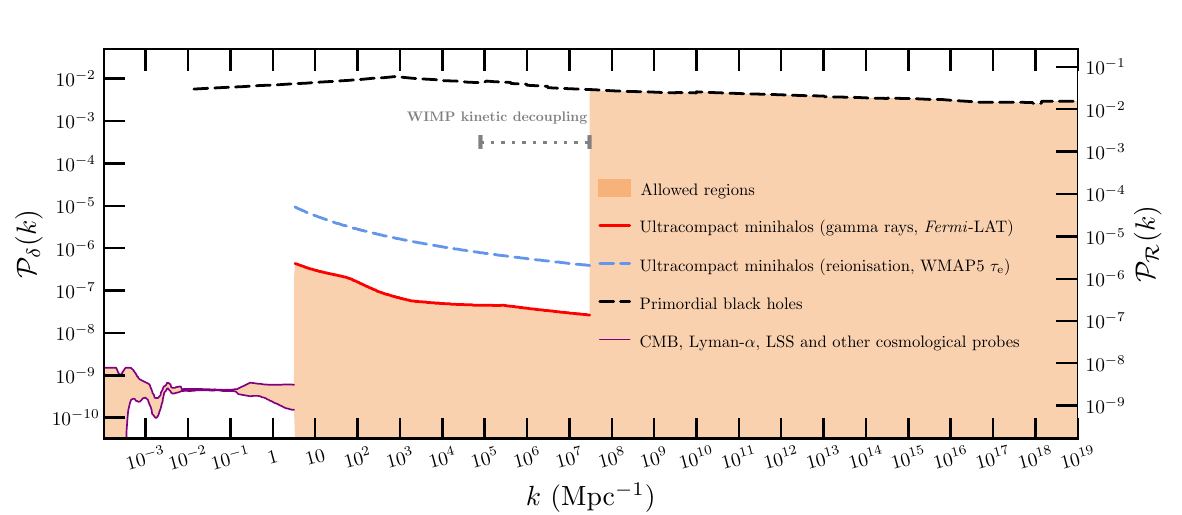}
\caption{Constraints on the allowed amplitude of primordial density (curvature) perturbations $\mathcal{P}_\delta$ ($\mathcal{P}_\mathcal{R}$) at all scales.  Here we give the combined best measurements of the power spectrum on large scales from the CMB, large scale structure, Lyman-$\alpha$ observations and other cosmological probes \cite{Nicholson:2009pi,Nicholson:2009zj,Bird:2010mp}.  We also plot upper limits from gamma-ray and reionisation/CMB searches for UCMHs, and primordial black holes \cite{JGM09}. For ease of reference, we also show the range of possible DM kinetic decoupling scales for some indicative WIMPs \cite{Bringmann:2009vf}; for a particle model with a kinetic decoupling scale  $k_\mathrm{KD}$, limits do not apply at $k>k_\mathrm{KD}$.  Note that for modes entering the horizon during matter domination, $\mathcal{P}_\delta$ (but not $\mathcal{P}_\mathcal{R}$) should be multiplied by a further factor of 0.81.}
\label{fig:all_constraints}
\end{figure*}

\subsection{Constraints from UCMHs}
\label{sec:UCMHs_constraints}

Using the limits on the mass variance of Gaussian perturbations presented in Fig.~\ref{fig:sigH_constraint}, we can now begin to constrain the primordial power spectrum.  This requires an explicit relationship between $\sigma^2_{\chi, \mathrm{H}}$ and the power spectrum itself.  We consider three broad models for the spectrum of perturbations: 
\begin{enumerate}
\item A scale-free spectrum $\mathcal{P}(k)\propto k^{n-1}$ parameterised by a spectral index $n$.
\item A scale-free spectrum with a step, resulting in increased power on small scales.  This spectrum is parameterised by the spectral index $n$, as well as the height $p$  and location $k_\mathrm{s}$ of the step:
\bea
 \mathcal{P}(k)&\propto& k^{n-1}
 \times 
\left\{ \begin{array}{ll} 
                1      & \mathrm{for}~k<k_s \\ 
             p^2 &  \mathrm{for}~k\geq k_s 
  \end{array} \right.\,,
\eea
\item A non-parametric generalised spectrum, assumed to be \emph{locally} scale-free in $k$-space, i.e.~$\mathcal{P}(k)\propto k^{n-1}$, but in general allowing for a different normalisation for very different values of $k$.
\end{enumerate}
In the first two cases, we normalize our spectra to the WMAP data in the same way as done in Ref.~\cite{Liddle:2006ev} (which is fully consistent with, and actually only marginally improves, the normalization measured by COBE \cite{cobe}).
For a more detailed description, and a derivation of the relationships between $\sigma^2_{\chi, \mathrm{H}}$ and the parameters (or running amplitude, in the generalised case) of each spectrum,  see Appendix \ref{app:sh}.  

Using these expressions, we translate the upper limits on $\sigma^2_{\chi, \mathrm{H}}$ in Fig.~\ref{fig:sigH_constraint} into upper limits on the parameters of each spectrum.  In Fig.~\ref{fig:n_constraint}, we give limits on the spectral index $n$ of the scale-free spectrum, derived from UCMHs of different masses, corresponding to different values of $k$.  Using such a spectrum, the most relevant limit is essentially the lowest one obtained at any scale.  Using this criterion, our limits on the UCMH abundance constrain the spectral index to be $n<1.17$.

In the case where the spectrum contains a step, we derive upper limits on the step size $p$ as a function of its location $k_\mathrm{s}$ (Fig.~\ref{fig:p_constraint}, left panel), assuming the spectral index $n=0.968$ observed on large scales by WMAP7 \cite{wmap}.  We also show how these limits change if $n$ is allowed to vary within the $1\sigma$ range of the WMAP7 measurements, plotting the resultant regions as shaded bands.  Depending upon the true value of $n$, the non-observation of UCMHs limits the size of a step in the power spectrum at any scale larger than $k=3\times10^7$\,Mpc$^{-1}$ to be less than a factor of 10--12, quite a severe constraint.

For each value of $k_\mathrm{s}$, the limit on $p$ is based on a particular scale $k$ (and therefore one specific $M^0_\mathrm{UCMH}$).  For each $k_\mathrm{s}$, we chose the optimal value of $k$ by minimising the resultant $p_\mathrm{max}$ over all $k$ using Brent's Algorithm.  As is to be expected, optimal limits always come from length scales smaller than the scale of the break, as this is the region affected by the step.  Because the limits on $\sigma^2_{\chi, \mathrm{H}}$ from gamma rays and reionisation are (albeit only approximately, in the case of gamma rays) monotonically decreasing functions of $k$, the optimal $k$ for placing limits on $p$ is always the smallest scale considered valid.  In our case, the dominating scale is $k=3\times10^7$\,Mpc$^{-1}$, which corresponds to our minimal assumed free-streaming mass cutoff, $M_i=5\times10^{-12}\,M_\odot$ ($M^0_\mathrm{UCMH}\sim10^{-9}\,M_\odot$).  The limits in the left panel of Fig.~\ref{fig:p_constraint} are therefore entirely dominated by the free-streaming scale, and should be treated with caution, as they will weaken for DM models that lead to larger minimal halo masses.  In the right panel of Fig.~\ref{fig:p_constraint}, we give an example of how the limits on $p$ for a spectrum with a step at $k_s=10^4$\,Mpc$^{-1}$ weaken if the kinetic decoupling scale is varied.  Here we consider the range $M_i=3\times10^{-4}-5\times10^{-12}\,M_\odot$, corresponding to allowed values within an indicative set of minimal supersymmetric standard model (MSSM) benchmark points \cite{Bringmann:2009vf}.  As expected, larger minihalo mass cutoffs weaken the corresponding limits on the size of a step in the primordial spectrum; the exception to this behaviour occurs for gamma-ray limits below $\sim$$3\times10^5$\,Mpc$^{-1}$, where the limits on $\sigma^2_{\chi, \mathrm{H}}$ in Fig.~\ref{fig:sigH_constraint} are not monotonically decreasing with $k$.

In the left panel of Fig.~\ref{fig:PR_constraint} we give limits upon the amplitude of curvature perturbations $\mathcal{P}_\mathcal{R}$ at small scales, using a non-parametric description of the spectrum.  The corresponding limits on the amplitude of physical density perturbations $\mathcal{P}_\delta$ can be obtained by simply multiplying the curvature perturbation limit by a factor of 0.191 (Eq.~\ref{pdelta_conversion}).  For comparison, we also show the rather different limits on curvature perturbations presented in Ref.~\cite{JG10}, derived from \textit{Fermi} searches for DM annihilation.  We attribute the differences in the two limits to the scale-dependent calculation of $\delta_\mathrm{min}$ we perform here, our improved statistical treatment of the limits coming from \textit{Fermi}, the inclusion of diffuse gamma-ray limits, and our corrected calculation of the mass variance of perturbations relative to Ref.~\cite{JG10}.

We also show in the right panel of Fig.~\ref{fig:PR_constraint} the dependence of our limits upon the WIMP mass, and the latest allowed redshift of UCMH collapse.  As argued earlier, our choices clearly are indeed very conservative.  Smaller WIMP masses and, to a much larger extent, later redshifts of collapse result in significantly strengthened limits -- simply because in the latter case much smaller initial density perturbations would have time to collapse, see Appendix \ref{app:dmin} and in particular Fig.~\ref{fig:deltamin}.
 The minimum redshift of collapse could actually quite defensively be made somewhat smaller, without invalidating our earlier arguments about the UCMH survival probability and violation of the radial infall approximation.  In the interests of producing as robust limits as possible, we use $z_c=1000$ as our canonical value, but it is worth noting that the substantial improvement in limits with smaller $z_c$ bodes well for the potential for future detection of UCMHs. 

\subsection{Comparison with existing constraints}

On large scales, the primordial power spectrum of density fluctuations has been measured with high precision mainly by CMB experiments (as the most powerful single source of information about the primordial fluctuations), large scale galaxy surveys (large scale structure; LSS) and weak gravitational lensing observations. However, the scales probed by such cosmological measurements constitute only a relatively small part of the entire spectrum, namely scales between $k \sim 10^{-4}$ Mpc$^{-1}$ and $k \sim 1$ Mpc$^{-1}$. This range has been extended to smaller scales ($k \sim 3$ Mpc$^{-1}$) by other astrophysical measurements that probe later epochs in the evolution of the Universe, such as the Lyman-$\alpha$ forest. For the rest of the spectrum, i.e. on scales smaller than $k \sim 3$ Mpc$^{-1}$, constraints have been provided mainly from non-observation of PBHs. Although PBHs constrain the power spectrum over a very wide range of small scales (from $k \sim 10^{-2}$ Mpc$^{-1}$ to $k \sim 10^{23}$ Mpc$^{-1}$), the constraints are different from those provided by the aforementioned cosmological probes in that 1) they are upper limits rather than positive measurements and 2) are much weaker (the upper limits are many orders of magnitude larger than the cosmological constraints on large scales).

A number of different techniques have been used to constrain the primordial power spectrum and its properties with cosmological observations. These include top-down approaches, where specific theoretical models of the primordial fluctuations or the inflaton potential are fit to the data, and bottom-up attempts to reconstruct the shape of the spectrum from data with no such assumptions. Such reconstruction techniques consist of simple binning techniques~\cite{Hannestad:2000pm,Mukherjee:2003cz,Bridle:2003sa,Mukherjee:2003ag,Hannestad:2003zs,Bridges:2005br,Spergel:2006hy,Bridges:2006zm,Hlozek:2011pc,Guo:2011re}, principal component analysis \cite{Leach:2005av}, methods of direct inversion~\cite{Wang:2000js,Matsumiya:2001xj,Kogo:2003yb,Shafieloo:2003gf,TocchiniValentini:2004ht,Kogo:2004vt,TocchiniValentini:2005ja,Shafieloo:2006hs,Shafieloo:2007tk,Nagata:2008tk,Nagata:2008zj,Ichiki:2009zz,Nicholson:2009pi,Nicholson:2009zj,Hamann:2009bz} and minimally-parametric reconstructions based on cross-validation technique \cite{Peiris:2009wp,Bird:2010mp}. Results of these reconstruction procedures, although consistent with each other, are in general not identical even in cases where the same observational datasets are employed. 

One of our objectives in this paper is to present a comprehensive compilation of the best constraints on the power spectrum at all scales, including those from UCMHs (as discussed in Sec.~\ref{sec:UCMHs_constraints}). We therefore select the strongest available constraints provided by the latest analyses of different cosmological data; these come from Refs.~\cite{Nicholson:2009pi,Nicholson:2009zj,Bird:2010mp}, and draw primarily upon CMB, LSS and Lyman-$\alpha$ data, though some small additional constraining power is derived from measurements of primordial nucleosynthesis, supernovae and the Hubble constant (refer to Ref.~\cite{Nicholson:2009zj} for details).  We combine these constraints into a single $1\sigma$ band (comprised of the best available upper and lower limits at each $k$). We plot the resultant constraint band in Fig.~\ref{fig:all_constraints}, along with our own results on small scales from reionisation and gamma-ray searches for individual UCMHs and Galactic diffuse DM sources.

We also show the current strongest upper limits on the power spectrum derived from PBHs~\cite{JGM09}, based on their present-day gravitational influence (for $k\lesssim10^{16}\,{\rm Mpc}^{-1}$) and  the products of their conjectured \cite{Hawking:1974sw} evaporation (for $k\gtrsim10^{16}\,{\rm Mpc}^{-1}$).  We do not show limits above $k\gtrsim10^{19}\,{\rm Mpc}^{-1}$, as such constraints rely on model-dependent assumptions about new (quasi-)stable elementary particles that often appear in extensions of the standard model.   At even smaller scales, $l_{\rm Pl}^{-1}\gtrsim k\gtrsim10^{21}\,{\rm Mpc}^{-1}$, the situation is even more uncertain: one must assume that the evaporation leaves a hypothetical Planck-size relic in order to place any further limits.  

We do not make any attempt to harmonise the CLs with which we state limits from different sources, as we do not have access to the full likelihood functions for any of the reported results. 
Let us also repeat a general word of caution for these kinds of limits: even though we present them here in a model-independent way, such limits always depend, to a certain extent, on the assumed \textit{spectrum} of the density perturbations.  This should be kept in mind when comparing  predictions from, e.g., inflationary models to what is shown in Fig.~\ref{fig:all_constraints} (see Appendix \ref{app:sh} for how to treat spectra that deviate from the \textit{locally} scale-free spectrum that we assumed here).

\section{Conclusions}
\label{sec:conclusions}

In $\Lambda$CDM cosmology, the first gravitationally bound objects typically form at redshifts considerably smaller than $z\sim100$. Very rare fluctuations, on the other hand, would collapse as early as the first stages of matter domination, forming UCMHs. By the time standard structure formation starts, these objects would have already developed a dense and highly concentrated core, which would be essentially unaffected by tidal interactions during subsequent cosmological evolution.

Arguably, the most compelling class of candidates for the nature of DM are WIMPs.
Because UCMHs have survival probabilities close to unity, the non-observation of individual gamma-ray sources by \textit{Fermi}, as well as measurements of the diffuse gamma-ray background, can then be used to place rather strong limits on the power spectrum of primordial density perturbations.  Potential changes in the reionisation history of the Universe, which would leave a visible imprint in the cosmic microwave background, can also provide some constraints.

We have provided and discussed in detail all necessary formulae to calculate these constraints for any functional form of the primordial spectrum of density fluctuations. As a possible application, we have constrained the spectral index of an assumed featureless power-law spectrum to be $n\lesssim1.17$. Since large-scale observations actually put much stronger limits on the spectral index, we have also considered the case of $n=0.968\pm0.012$, as obtained by WMAP observations, and constrained the allowed additional power below some small scale $k_s$ to be at most a factor of $\sim$10--12 (assuming a step-like enhancement in the spectrum). As a third example, we have obtained quasi-model-independent limits, of the order of $\mathcal{P}_\mathcal{R}\lesssim10^{-6}$, on perturbation spectra  that can at least locally be well described by a power law. We would like to stress, however, that it is intrinsically impossible to constrain primordial density fluctuations in a completely model-independent way; one thus has to re-derive such limits for any particular model of, e.g., inflation which produces a spectrum that does not fall into one of these classes. Here, we have provided all the necessary tools to do so.

We have mentioned that present gravitational lensing data  cannot be used to constrain the abundance of UCMHs -- essentially because they are simply not point-like enough, even in view of their highly dense and concentrated cores. Future missions making use of the light-curve shape in lensing events, however, are likely to probe or constrain their existence. This would be quite remarkable as it would allow us to put limits on the power spectrum without relying on the WIMP hypothesis for DM. Most of our formalism is readily extended, or can in fact be directly applied to, such constraints arising from gravitational microlensing.

Finally, we have compiled an extensive list of the most stringent limits on $\mathcal{P}_\mathcal{R}(k)$ that currently exist in the literature for the whole range of accessible scales, from the horizon size today down to scales some 23 orders of magnitude smaller. Direct and indirect observations of the matter distribution on large scales -- in particular galaxy surveys and CMB observations -- constrain the power spectrum to be $\mathcal{P}_\mathcal{R}(k)\sim 2\times10^{-9}$ on scales larger than about $1\,$Mpc. On sub-Mpc scales, on the other hand, only upper limits exist.
From the non-observation of PBH-related effects, one can infer $\mathcal{P}_\mathcal{R}\lesssim10^{-2}-10^{-1}$ on all scales that we consider here. UCMHs are much more abundant and thus result in considerably stronger constraints, $\mathcal{P}_\mathcal{R}\lesssim10^{-6}$, down to the smallest scale at which DM is expected to cluster (this depends on the nature of the DM; for typical WIMPs like neutralino DM, e.g., it falls into the range $k_{\rm max}^\chi\sim 8\times10^4-3\times10^7\,{\rm Mpc}^{-1}$).

It is worth recalling that the observational evidence for a simple, nearly Harrison-Zel'dovich spectrum of density fluctuations is obtained by probing a relatively small range of rather large scales. The limits we have provided here will thus be very useful in constraining any model of e.g.~inflation, or phase transitions in the early Universe, that predicts deviations from the most simple case and which would result in more power on small scales.

\smallskip
\acknowledgments
We warmly thank Sofia Sivertsson for her input during the early stages of this work, and Adrienne Erickcek for very useful comments on the manuscript.  We are also grateful to Venya Berezinsky and Alexander Westphal for enlightening discussions about the tidal destruction of dense DM clumps and the spectrum of density fluctuations expected from inflation, respectively.  We thank Dong Zhang for illuminating comments regarding the reionisation limits.
T.B. acknowledges support from the German Research Foundation (DFG) through the Emmy Noether  grant BR 3954/1-1.  P.S. thanks the Astroparticle Group of the II. Institute for Theoretical Physics at the University of Hamburg for their hospitality, with support from the LEXI initiative Hamburg, whilst this paper was being conceived, and is supported by the Lorne Trottier Chair in Astrophysics and an Institute for Particle Physics Theory Fellowship. Y.A. thanks the Swedish Research Council (VR) for financial support. Y.A. was partially supported by the European Research Council (ERC) Starting Grant StG2010-257080.

\appendix

\section{Minimal density contrast for UCMH formation}
\label{app:dmin}

In their original paper, Ricotti \& Gould \cite{Ricotti:2009bs} used an order-of-magnitude estimate to argue that a region of co-moving size $R$ should have an average over-density of $\delta\gtrsim\delta^{\rm min}=10^{-3}$, at the time of horizon crossing,\footnote{
In Fourier space, the time $t_k$ of horizon crossing is conventionally defined as $\left.aH\right|_{t_k}=k$, where $k$ is the wavenumber of the perturbation. In real space, we define it as $aR=H^{-1}$; this amounts to saying that we associate a wavenumber $k_R\equiv R^{-1}$ to a perturbation of co-moving size $R$.}
in order to gravitationally collapse to a UCMH before a redshift of $z\sim1000$. This is also the value that has been used in subsequent work, see e.g.~Refs.~\cite{SS09,JG10,Zhang:2010cj}. However, since $\delta^{\rm min}$ enters exponentially in the expression for the abundance of UCMHs, it is important to have a more accurate estimate of this quantity. Note, in particular, that one should expect it to be scale-dependent rather than just a constant since DM perturbations continue to grow after they enter into the horizon (even though this growth is only logarithmic for times long before matter-radiation equality \cite{Meszaros:1974tb}).

To derive the minimal density contrast required for UCMH formation, 
we will in the following rely on a simplified description for the collapse of over-dense regions that was originally introduced in \cite{peebles67} and is now widely being used. The basic  approximation is to restrict oneself to the case of spherical regions of uniform density $\rho(t)=\bar\rho(t)+\delta\rho(t)$ that are embedded in a background with density $\bar\rho(t)$. During matter domination, the full (non-linear) evolution of the over-density can in that case be solved exactly; $\rho(t)$ starts off by decreasing more slowly than the background density $\bar\rho(t)$, reaches a minimum after a finite time $t_c/2$ and then increases until it becomes infinitely large at $t=t_c$. In this approach, the time $t_c$ is thus identified as the time where the region has fully collapsed -- and which at the same time indicates the breakdown of this simplified description (in a more realistic description, dynamic relaxation and angular momentum conservation would of course prevent the collapse to a point-like, singular object). In the \emph{linear} theory, on the other hand, the over-density would by that time have grown to
\be
\label{deltac}
\delta_c\equiv\frac{\delta\rho(t_c)_{\rm lin}}{\bar \rho(t_c)}=\frac{3}{5}\left(\frac{3\pi}{2}\right)^{2/3}\approx1.686\,.
\ee
This relation thus allows us to use the linear theory of perturbations to calculate the time of collapse (at least during matter domination) -- which is quite remarkable since the perturbations of course enter the non-linear regime already quite some time before the actual collapse.

Unfortunately, even in linear theory the system of equations governing the evolution of density contrasts around the time of matter-radiation equality is rather complicated and cannot be solved analytically. However, a very accurate fit to a numerical solution for the density contrast in DM fluctuations for $t>t_{\rm eq}$ can be found, e.g., in Ref.~\cite{cosmology:Weinberg}:
\be
\label{dchifull}
 \delta_\chi(k,t)=\frac{9k^2t^2}{10\,a^2}\mathcal{T}(\kappa)\mathcal{R}^0(k)\,,
\ee
where $k$ is the co-moving wave number of the perturbation and $\kappa$ its rescaled, dimension-less version:
\be
 \kappa\equiv\frac{\sqrt{2}k}{a_{\rm eq}H_{\rm eq}}=\frac{k\sqrt{\Omega_r}}{H_0\Omega_m}\,.
\ee  
The fitting function $\mathcal{T}(\kappa)$  is given by
\bea
  \mathcal{T}(\kappa) &\simeq&
\frac{\ln[1+(0.124\kappa)^2]}{(0.124\kappa)^2}   \\
& \times&\left[\frac{1+(1.257\kappa)^2+(0.4452\kappa)^4+(0.2197\kappa)^6}{1+(1.606\kappa)^2+(0.8568\kappa)^4+(0.3927\kappa)^6}\right]^{1/2}.\nonumber
\eea
The above expression for $\delta_\chi$ is given in synchronous gauge, but since we are interested in scales that are much smaller than the horizon at the time of collapse, the choice of gauge does not actually matter. The normalisation is chosen such that, for adiabatic fluctuations, $\mathcal{R}^0$ gives the value of the initial curvature perturbation (which is time-independent on scales much larger than the horizon).

By equating Eqs.~(\ref{deltac}) and (\ref{dchifull}), we can thus derive the minimal value of $\mathcal{R}^0$ that is required so that the perturbation in the DM component collapses before a given redshift $z_\mathrm{c}<z_{\rm eq}$:
\be
\label{Rmin}
\mathcal{R}^0_\mathrm{min}(k) = \left.\frac{a^2}{t^2}\right|_{z=z_\mathrm{c}} \frac23\left(\frac{3\pi}{2}\right)^\frac23\frac{1}{\mathcal{T}(\kappa)k^2}\,.
\ee
While $\mathcal{R}^0$ is the actual fundamental physical quantity describing the strength of adiabatic fluctuations -- predicted, e.g., in theories of inflation -- one may instead also consider the somewhat more intuitive value of the DM density contrast at the time a fluctuation enters into the horizon -- which in our case would be during the radiation-dominated era. In contrast to $\mathcal{R}^0$, however, the density contrast is a gauge-dependent quantity  (which becomes numerically quite relevant for scales $k\lesssim aH$). In the following, we will choose the \emph{co-moving (or ``total matter'') gauge}, where the rest frame is that of the total energy density fluctuations; this is the gauge that corresponds to the initial conditions adopted in the treatment of the collapse outlined above (and also is typically used, e.g., for the calculation of PBH formation). 

During radiation domination, the evolution of all perturbed quantities like $\delta_\chi$ can be solved analytically. To convert the results in synchronous gauge given in Ref.~\cite{cosmology:Weinberg} to the total matter frame, one has to perform a gauge transformation under which
\be
 \delta\rho^{\rm (S)} \rightarrow  \delta\rho^{\rm (T)} =\delta\rho^{\rm (S)} + \dot{\bar\rho}\, \delta u^{\rm (S)}, 
\ee
where $\dot{\bar\rho}$ is the time derivative of the mean density and $\delta u^{\rm (S)}$ is the scalar part of the velocity components in the stress-energy tensor in synchronous gauge.\footnote{
This transformation behaviour follows from the fact that both synchronous and total matter gauge have vanishing metric components $g_{0i}$ and that in co-moving gauge we have $\delta u=0$. Note that the latter condition follows from $\delta u$ being defined as part of the stress-energy tensor (as in Ref.~\cite{cosmology:Weinberg}) -- which is a somewhat different definition compared to other examples from the literature \cite{altdu}; as a result, the change of velocity perturbations under gauge transformations does not take the same form as in these references, either. Of course, the final results are unaffected.
}
The result can be written as
\be
\label{deltamattergauge}
\delta_i(k,t)=N_i\theta^2T_i(\theta)\mathcal{R}^0(k)\,,
\ee
where 
\be
\label{theta}
 \theta=\frac{2kt}{\sqrt{3}a}=\frac{1}{\sqrt{3}}\frac{k}{aH}
\ee
and $i=\chi,r$ for the DM and radiation component, respectively. For \emph{adiabatic} fluctuations we  have $N_r=4/3$ and $N_\chi=1$. 
The \emph{transfer functions} introduced here satisfy $T_i(0)=1$ and are given by
\bea
\label{Tr}
 T_\mathrm{r}(\theta)&=& \frac{3}{\theta} j_1(\theta)\,,\\
\label{Tx}
 T_\chi(\theta)&=&   \frac{6}{\theta^2} \left[\ln(\theta) +\gamma_{\rm E}-\frac{1}{2}-{\rm Ci}(\theta)+\frac{1}{2}j_0(\theta)\right],\hspace{8mm}
\eea
where $\gamma_{\rm E}$ is the Euler-Mascheroni constant, $\rm Ci$ the cosine integral function and $j_{0,1}$ are spherical Bessel functions of the first kind.  Eq.~(\ref{Tx}) is somewhat painful to implement numerically in certain cases; a good solution is to modify an existing code for computing ${\rm Ci}(x)$ (as found in e.g.~\cite{NR}) to instead return $x^{-2}[{\rm Ci}(x)-\ln x-\gamma_{\rm E}]$.

\begin{figure}[t]
\includegraphics[width=\columnwidth, trim = 0 20 0 0, clip=true]{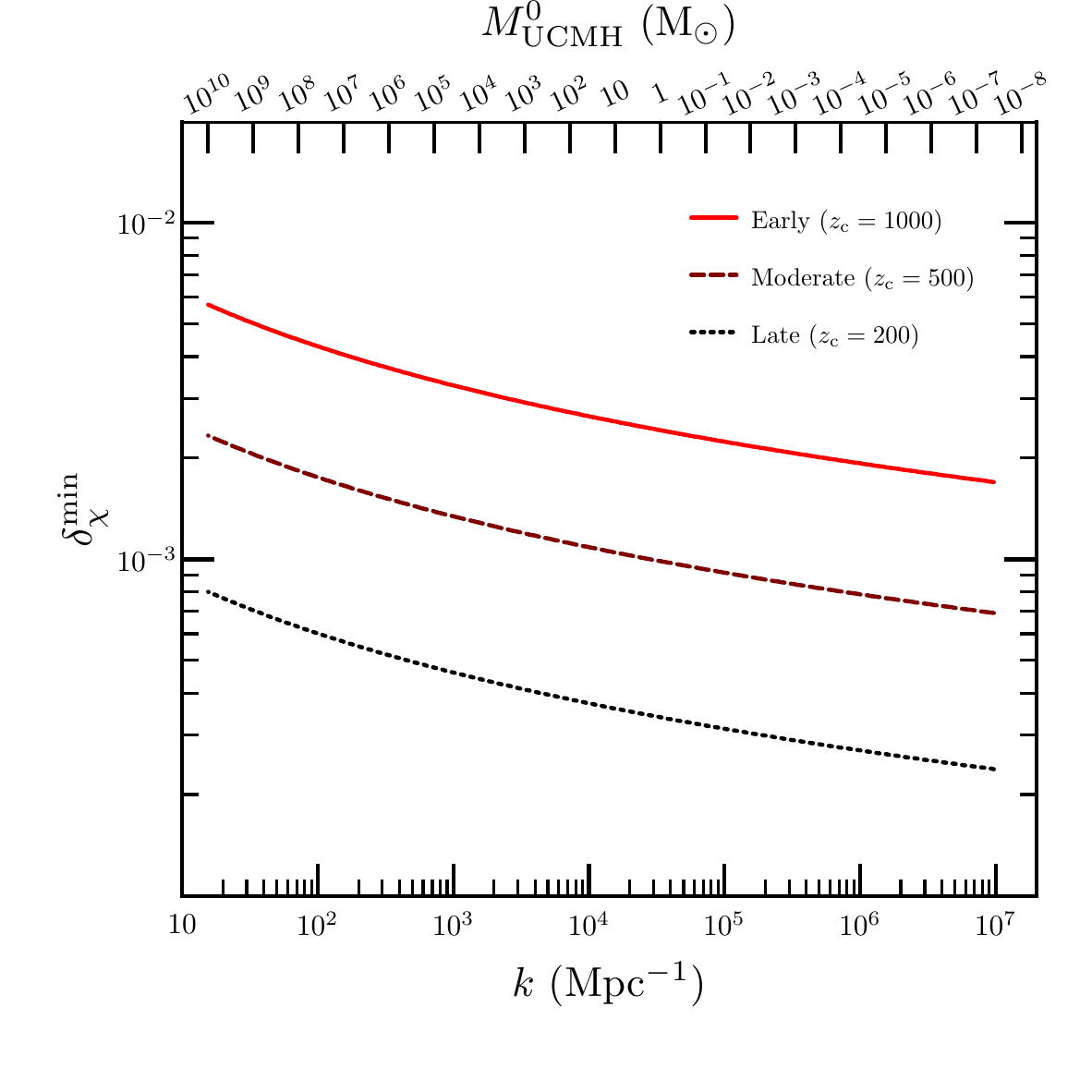}
\caption{Minimal density contrast in the DM component, at the time of horizon entry $(k=aH)$, required to form a UCMH before redshift $z_c=1000,\,500,$ or $200$.}
\label{fig:deltamin}
\end{figure}

We have now collected all the pieces needed to express the minimal density contrast in the DM component, at the time of horizon entry $t_k$ of a scale $k$, for a perturbation to collapse and form a UCMH before a redshift $z_\mathrm{c}$.  Combining Eqs.~(\ref{Rmin}) and (\ref{deltamattergauge}), we have
\be
 \delta_\chi^{\rm min}(k,t_k)=\left.\frac{a^2}{t^2}\right|_{z=z_\mathrm{c}}\frac{2}{9}\left(\frac{3\pi}{2}\right)^{2/3}\frac{T_\chi(\theta=1/\sqrt3)}{k^2\mathcal{T}(\kappa)}\,.
\ee
This function is plotted in Fig.~\ref{fig:deltamin} for a few selected values of $z_c$. Note that after decoupling at $z_{\rm dec}\sim1000$, baryons would also start to gravitationally collapse and thus significantly contribute to the over-density -- which has not been taken into account here. In that sense, our estimate for $\delta_\chi^{\rm min}$ is rather conservative for redshifts $z\ll z_{\rm dec}$.

\section{Correct normalisation of power spectrum and mass variance}
\label{app:sh}

In this Appendix, we review in detail how to express the mass variance, i.e.~the r.m.s.~over-density in a given region of space, in terms of the super-horizon spectrum of density or curvature fluctuations provided by inflation. 

Assuming Gaussian statistics for the primordial density fluctuations,  the probability (density) to find an average density contrast $\delta$ in a spherical region of size $R$ is given by
\be
p_R(\delta)=\frac{1}{\sqrt{2\pi}\sigma(R)}\exp\left[-\frac{\delta^2}{2\sigma^2(R)}\right]\,,
\ee 
where the mass variance $\sigma(R)$ is computed by convolving the power spectrum with a top-hat window function:
\be
\label{sig}
\sigma^2(R)=\int_0^\infty W^2_{\rm TH}(kR)\mathcal{P}_\delta(k)\frac{dk}{k}.
\ee
In the above expression, $\mathcal{P}_\delta(k)$ is defined by 
\be
\label{pdelta}
\langle\delta_\mathbf{k}\delta_\mathbf{k'}^*\rangle\equiv\frac{2\pi^2}{k^3}
\mathcal{P}_\delta(k)\,\delta(\mathbf{k}-\mathbf{k'})\,
\ee
and $W_{\rm TH}(x)=3j_1(x)/x=3\,x^{-3}\left(\sin x - x \cos x\right)$ denotes the Fourier transform of the 3D top-hat window function, with $x\equiv kR$. 

The above description is somewhat complicated by the fact that density perturbations evolve with the expansion of the Universe, which means that also the power spectrum is time-dependent. In total matter (as well as in synchronous) gauge, the quantity\footnote{
The quantity $\delta_\mathrm{H}$ is directly related to what is also known as the peculiar gravitational potential $\Phi$, $\langle\Phi^2\rangle=({9\pi^2}/{2k^3}) \delta_\mathrm{H}^2$.
}
\be
\label{deltaHsq}
\delta_\mathrm{H}^2(k,t)\equiv\frac{(aH)^4}{k^4}\mathcal{P}_\delta(k)\,,
\ee
however, is time-independent on super-horizon scales ($k\ll aH$), with a numerical value that is very close to the value at the time $t_k$ when mode $k$ crosses the horizon.
It is thus illustrative to separate the power spectrum into a part that describes the primordial fluctuation spectrum on super-horizon scales, as provided by inflation, and a part that encodes the evolution of the perturbations (mostly after they enter the horizon). This is usually done by introducing a transfer function
\be
T^2(k,t)\equiv \delta_\mathrm{H}^2(k,t)/\delta_\mathrm{H}^2(k,t_i)\,,
\ee
which satisfies $T(k\!\rightarrow\!0,t)\rightarrow1$. In the above definition, $t_i$ denotes a time before the entrance of any scale $k$ into the horizon, $t_i\!<\!t_k\, \forall k$, so it should be taken to correspond to the time at the end of inflation.  Note that the time-dependence of the transfer function only enters as the ratio of $aH/k$, so $T$ can also be written as a function of one variable only.  For e.g.~radiation domination, explicit expressions are given in Eqs.~(\ref{theta}--\ref{Tx}). The relation between $\delta_\mathrm{H}$ and the initial spectrum of (adiabatic) curvature perturbations during radiation domination is
\be
\label{curvaturespectrum}
 \mathcal{P}_\mathcal{R}(k)=\left(\frac{3}{2}\right)^4\delta_\mathrm{H}^2(k,t_i),
\ee
as can be explicitly verified with the help of Eqs.~(\ref{theta}--\ref{Tr}).
During matter domination, the right hand side of this expression has to be multiplied by $(10/9)^2$ \cite{Polarski:1992dq}.

As explained in Appendix \ref{app:dmin}, the condition for UCMH formation can be stated in terms of the DM density contrast at horizon crossing of a given scale. We should therefore evaluate the mass variance in Eq.~(\ref{sig}) at the time when $k_R\equiv1/R=aH$. With the shorthand notation
\be
\delta_\mathrm{H}(t_k)\equiv\delta_\mathrm{H}(k,t_k)\,,
\ee
this can be conveniently expressed as  \cite{pbh1,pbh2}
\be
\sigma^2_\mathrm{H}(R)\equiv \left.\sigma^2\left(R\right)\right|_{t=t_{k_{R}}}\!\!=\alpha^2(k_R)\,\delta_\mathrm{H}^2(t_{k_R})\,,
\ee
where
\be
\label{defalpha}
 \alpha^2(k)= \delta_\mathrm{H}^{-2}(t_k)\int_0^\infty x^3\delta_\mathrm{H}^2(xk,t_i)T^2(xk,t_k)W^2_{\rm TH}(x) dx\,.\ee
The important point to note here \cite{pbh1,pbh2} is that the relation between mass variance $\sigma_\mathrm{H}$ and size of the perturbation at horizon crossing $\delta_\mathrm{H}$ depends on the scale and, in principle, on the full cosmological evolution between the end of inflation and $t_k$.

Unlike for the case of PBHs, only the density contrast in the DM component will grow and eventually collapse to a UCMH; in the above expressions, we should thus place an index $\chi$ where appropriate. Since $\delta$ and $\delta_\chi$ only differ by a constant factor on super-horizon scales (at least for curvature perturbations), the main difference between $\alpha(k)$ and $\alpha_\chi(k)$ is the transfer function that is used. This difference, however, is important because perturbations in the DM component behave completely differently from those in other components once they enter the horizon, cf.~Eqs.~(\ref{Tr}--\ref{Tx}).
The mass variance we are really interested in is therefore given by
\be
\label{sHx}
\sigma^2_{\chi,\mathrm{H}}(R)=\alpha_\chi^2(k_\mathrm{R})\,\delta_\mathrm{H}^2(t_{k_R})\,,
\ee
where $\delta_\mathrm{H}^2(t_{k_R})$ refers to the \textit{total} energy fluctuations because we later want to normalise it to the present-day density contrast observed in the CMB.  For $k\gg k_{\rm eq}=\sqrt{2}H_0\Omega_m/\sqrt{\Omega_r}$, we have
\be
\label{defalphachi}
 \alpha_\chi^2(k)= \frac{9}{16}\int_0^\infty \!\!dx\,x^3W^2_{\rm TH}(x)\frac{\mathcal{P}_\mathcal{R}(xk)} {\mathcal{P}_\mathcal{R}(k)}  \frac{T_\chi^2\left(\theta=x/\sqrt{3}\right)} {T_\mathrm{r}^2\left(\theta=1/\sqrt{3}\right)}\,.
\ee

\subsection{Scale-free spectrum}
\label{scalefree_norm}

It is often assumed that the spectrum on super-horizon scales is of a scale-free form, implying also that
\be 
\label{scalefree}
\delta^2_\mathrm{H}(t_k)\propto k^{n-1}
\ee 
during any epoch where the equation of state does not change. In this case, Eq.~(\ref{defalphachi}) simplifies to
\be
\label{alphan}
 \alpha_\chi^2(n)= \frac{9}{16}\int_0^\infty \!\!dx\,x^{n+2} W^2_{\rm TH}(x)  \frac{T_\chi^2\left(x/\sqrt{3}\right)} {T_\mathrm{r}^2\left(1/\sqrt{3}\right)}\,.
\ee

For a spectrum like in Eq.~(\ref {scalefree}) and a flat cosmology, the WMAP normalisation 
at the scale $k_{\rm WMAP}=0.05\,$Mpc$^{-1}$ is given by \cite{Liddle:2006ev} 
\be
\delta_\mathrm{H} = 1.927 \times10^{-5}\frac{\exp\left[(1-n)(-1.24+1.04r)\right]}{\sqrt{1+0.53r}}\,,
\ee
where $r$ is the tensor-to-scalar ratio. Assuming no gravitational waves (i.e.~$r=0$) and taking into account that  the normalisation of $\mathcal{R}$
leads to an effective suppression of $\delta_\mathrm{H}$ by a factor of $9/10$ during the transition to the matter dominated phase  \cite{Polarski:1992dq}, we thus have for $t_k\ll t_{\rm eq}$:
\bea
\label{dhn}
\delta_\mathrm{H}^2(t_k)&=&\left(\frac{10}{9}\right)^2\, \delta^2_\mathrm{H}(t_{k_{\rm WMAP}}) \, \left(\frac{k}{k_{\rm WMAP}}\right)^{n-1}\\
&\simeq&4.58\times10^{-10}e^{2.48(n-1)}\left(\frac{k}{k_{\rm WMAP}}\right)^{n-1}\,.\nonumber
\eea
For non-zero values of $r$, in principle, this normalization changes slightly. From inflation in the slow-roll approximation, e.g., we expect $r\geq(8/3)(1-n)\simeq0.085$ (assuming the spectral index observed on large scales), corresponding to a very modest decrease of $\delta_\mathrm{H}$ by about 2\%. Even the largest currently allowed value of $r\sim0.3$ \cite{wmap} would decrease $\delta_\mathrm{H}$ by only about 6\%, leading to a corresponding weakening of our limits on the height $p$ of a step in the spectrum (see below) and barely affecting our limits on $n$.

We now have all the ingredients to calculate $\sigma_{\chi, H}$ as in Eq.~(\ref{sHx}).
Approximating the transition between radiation and matter domination to take place almost instantaneously, the last factor in the above expression can, furthermore, be expressed in terms of the total mass contained within the horizon as
\be
 \left(\frac{k}{k_0}\right)^{n-1}\approx \left[\frac {M_H(t_0)} {M_H(t_{\rm eq})}\right]^{(n-1)/3}\left[\frac{M_H(t_{\rm eq})} {M_H(t_k)}\right]^{(n-1)/2}\,,
\ee
where the first term on the right-hand side can also be written as $(1+z_\mathrm{eq})^{n-1}$.

\subsection{Spectrum with a superimposed step}
\label{withstep_norm}

The most simple phenomenological way to allow for more power below some small scale ($k\gg k_{\rm eq}$) is to generalise the scale-free case by introducing two parameters ($k_s, p$) that describe the location and height of a step in the primordial spectrum (see e.g.~\cite{pbh1,pbh2}):
\bea
 \delta^2_{{\rm H, step}}(t_k)&=&\delta^2_\mathrm{H}(t_k)
 \times 
\left\{ \begin{array}{ll} 
                1      & \mathrm{for}~k<k_s \\ 
             p^2 &  \mathrm{for}~k\geq k_s 
  \end{array} \right.\,,
\eea
where $\delta^2_\mathrm{H}(t_k)$ is given by Eq.~(\ref{dhn}).
We can then again use  Eq.~(\ref{sHx}) to calculate $\sigma_{\chi, \mathrm{H}}$ by using  
\bea
 \alpha_{\chi,{\rm step}}^2(k,k_s,p,n)=\alpha^2_\chi(n)+\frac{9(p^2-1)}{16}\hspace{15mm}\\
 \hspace{15mm}\times\int_{k_s/k}^\infty \!\!dx\,x^{n+2} W^2_{\rm TH}(x)  \frac{T_\chi^2\left(x/\sqrt{3}\right)} {T_\mathrm{r}^2\left(1/\sqrt{3}\right)}\,,\nonumber
\eea
with $\alpha_\chi(n)$ given by Eq.~(\ref {alphan}).
Note that a rather similar spectrum can arise if the inflaton potential has a jump in its first derivative \cite{Starobinsky:1992ts}.

\subsection{Generalised spectrum and curvature perturbations}
Instead of relating it to $\delta_\mathrm{H}$, we can of course also express the mass variance directly in terms of the curvature perturbation and write Eq.~(\ref{sHx}) as
\be
\label{sig2_gen}
\sigma^2_{\chi,\mathrm{H}}(R) = \frac19 \int_0^\infty x^3 W^2_{\rm TH}(x)\mathcal{P}_\mathcal{R}(x/R)T_\chi^2\left(x/\sqrt{3}\right)\,\mathrm{d}x.
\ee
This integral is strongly dominated by contributions around $x\sim1$ (i.e.~by wavenumbers $ k_\mathrm{R}=1/R$).  A general form for the spectrum of primordial curvature perturbations can be obtained by discarding the assumption of a globally scale-free spectrum, and instead adopting the weaker assumption of \emph{local} invariance only around the scale of interest $k_\mathrm{R}$ probed by (non-)observation of UCMHs of a certain mass (see e.g.~\cite{JG10}).  Given the dominance of $k_\mathrm{R}$ in the integral above, this can be achieved by taking a power-law form for $\mathcal{P}_\mathcal{R}(k)$, with the pivot point placed at $k_\mathrm{R}$,
\be
\mathcal{P}_\mathcal{R}(k) = \mathcal{P}_\mathcal{R}(k_\mathrm{R})\left(\frac{k}{k_\mathrm{R}}\right)^{n_\mathrm{R}(k_\mathrm{R})-1}.
\ee
Here $n_\mathrm{R}(k_\mathrm{R})$ is the \emph{local} slope of the power-law at $k_\mathrm{R}$.  For e.g.~slow-roll inflationary models, $0.9\lesssim n_\mathrm{R}\lesssim1.1$ is typical \emph{at scales measured by the CMB} \cite{inflation_review}; for the considerably smaller scales we are mainly interested in here, however, it is much more difficult to make general statements about the expected value of $n_R$. 
For  our constraints we use $n_\mathrm{R}=1$, for which evaluation of Eq.~(\ref{sig2_gen})  gives
\be
 \sigma^2_{\chi,{\rm{H}}}(R)/ \mathcal{P}_\mathcal{R}(k_\mathrm{R})=  0.907\,.
\ee
For comparison, changing $n_\mathrm{R}$ to $0$ ($2$) would result in $0.388$ ($2.91$) 
-- which should serve as a warning that it is in general not possible to translate bounds on $\sigma^2_{{\rm{H}}}$ into bounds on $ \mathcal{P}_\mathcal{R}$ in a completely model-independent way, i.e.~without assuming anything about the form of the spectrum.

Similarly, using Eqs.~(\ref{deltaHsq}--\ref{curvaturespectrum}) we can express the equivalent total density power spectrum in terms of the initial curvature spectrum as
\be
\label{pdelta_pr}
\mathcal{P}_\delta(k) = \left(\frac{2k}{3aH}\right)^4T^2_r(k,t)\mathcal{P}_\mathcal{R}(k),
\ee
which at the time of horizon entry gives 
\be
\label{pdelta_conversion}
\mathcal{P}_\delta(k)|_{t=t_k} = \left(\frac{2}{3}\right)^4T^2_r(\theta=1/\sqrt3)\mathcal{P}_\mathcal{R}(k)=0.191\,\mathcal{P}_\mathcal{R}(k).
\ee


\end{document}